\documentclass[journal=jacsat,manuscript=article]{achemso}

\usepackage[version=3]{mhchem} 
\usepackage{array}

\author{Mariia Stepanova}
\affiliation{Department of Electrical Engineering and Computer Science, University of California, Irvine, CA 92697, USA}

\author{Minh Ngo}
\affiliation{Department of Electrical Engineering and Computer Science, University of California, Irvine, CA 92697, USA}
\author{Mashnoon Alam Sakib}
\affiliation{Department of Electrical Engineering and Computer Science, University of California, Irvine, CA 92697, USA}
\author{Wills Harris}
\affiliation{Department of Physics, University of California, Irvine, CA 92697, USA}
\author{Joshua Bocanegra}
\affiliation{Department of Physics, University of California, Irvine, CA 92697, USA}
\author{Ruqian Wu}
\affiliation{Department of Physics, University of California, Irvine, CA 92697, USA}
\author{Kristie J. Koski}
\affiliation{Department of Chemistry, University of California, Davis, CA 95616, USA}
\author{Maxim R. Shcherbakov}
\affiliation{Department of Electrical Engineering and Computer Science, University of California, Irvine, CA 92697, USA}
\alsoaffiliation{Department of Materials Science and Engineering, University of California, Irvine, CA 92697, USA}
\alsoaffiliation{Beckman Laser Institute, University of California, Irvine, CA 92697, USA}
\email{maxim.shcherbakov@uci.edu}
\title[An \textsf{achemso} demo]
  {Highly Tunable Phonon Polaritons via Metal Intercalation}


\begin{document}







\begin{abstract}

Phonon polaritons in 
van der Waals crystals offer 
mid-infrared light confinement deep below the diffraction limit, 
making them promising 
for nanophotonics applications. However, the practical 
use of phonon polaritons 
remains limited, in part 
due to the lack of precise control over the phonon polariton dispersion, as 
crystal lattice vibrations are often inert to external stimuli. Here, we 
address this challenge 
by zerovalent metal intercalation of $\alpha$-MoO$_3$. 
 Photo-induced force microscopy shows that introducing tin 
 into the van der Waals gap modulates the phonon polariton dispersion by up to $38.5\pm±0.5\%$, which is the highest amount of tunability among non-mechanical modulation approaches, to the best of our knowledge. Intercalation with various metal species 
 preserves the phonon polariton lifetimes, while modulating the dielectric permittivity in agreement with the density functional theory and analytical calculations. 
Our results establish zerovalent metal intercalation as a practical route to reconfigurable mid-infrared nanophotonics
. 



\end{abstract}


Polaritons---hybrid quasiparticles formed by the strong coupling of light and matter\cite{Low2016}---hold promise for controlling the flow of light at the nanoscale \cite{Taboada2024}. 
In recent years, layered van der Waals (vdW) materials have become an attractive platform for studying and manipulating polaritons. 
\cite{Wu2022,Basov2016}. 
Phonon polaritons (PhPs) originate from photons coupled to 
optical phonons 
in polar dielectric crystals \cite{Foteinopoulou2019,Galiffi2023}. PhPs emerge in the infrared (IR) 
and demonstrate strong sub-wavelength field confinement,
 low propagation losses and long lifetime due to the absence of free charge carriers \cite{Hu2020,Ma2024}. 
The extraordinary properties of PhPs in vdW crystals can enable 
sub-diffraction focusing \cite{duan_planar_2021, zheng_controlling_2022, liang_manipulation_2024}, imaging \cite{li_hyperbolic_2015, duan_canalization-based_2025, dai_subdiffractional_2015}, and molecular detection \cite{autore_boron_2018,bareza_phonon-enhanced_2022}, making them highly attractive for nanophotonic applications. 

Control over fundamental characteristics, such as dispersion, is required to enable practical implementation of PhPs. 
During the past decades, the manipulation of the PhP dispersion has been demonstrated through a variety of approaches, such as changing the local dielectric environment \cite{zheng_tunable_2022, alvarez-perez_negative_2022,yu_hyperbolic_2023}, as well as substrate-mediated canalization \cite{duan_canalization-based_2025}, twist-induced topological transition \cite{hu_topological_2020}, edge-oriented steering \cite{dai_edge-oriented_2020}, and gate-tunability in heterostructures \cite{hu_gate-tunable_2023}. 
All these 
techniques rely on external 
stimuli, to which 
crystal lattice vibrations are often inert, limiting the dispersion tunability
. One way to overcome this challenge is by directly modifying 
crystal structure of the polaritonic material itself \cite{Teng2024,sakib2025vacancyengineeredphononpolaritonsvan}. 
Intercalation is one of the chemical approaches 
to altering the homogeneity of the host material through the inclusion of foreign atoms 
between the weakly bound layers of the vdW materials \cite{Koski2012,Reed2019,Reed2024,Reed2024+}. Early works have successfully demonstrated the controlled tunability of PhPs by intercalation in $\alpha$-MoO$_3$ \cite{wu_chemical_2020,zheng_highly_2018,wu_efficient_2021,zhao_ultralow-loss_2022} and $\alpha$-V$_2$O$_5$ \cite{taboada-gutierrez_broad_2020} vdW crystals. The hydrogen intercalation of $\alpha$-MoO$_3$ reversibly switches PhP \cite{wu_chemical_2020}, while sodium-intercalated $\alpha$-V$_2$O$_5$ enables broad spectral shift of PhPs Reststrahlen band (RB) \cite{taboada-gutierrez_broad_2020}.
 However, only a few intercalant materials have been attempted so far to demonstrate proof-of-concept for the manipulation of PhPs via intercalation, while a wide variety of metal intercalations can modulate atomic vibrations of vdW crystals, as 
 shown 
 by changes in Raman spectra \cite{zheng_highly_2018,https://doi.org/10.1002/aenm.201803137,doi:10.1021/acs.jpclett.7b03374, xu_electronic_2020}.  

In this work, we demonstrate the manipulation of PhP dispersion in $\alpha$-MoO$_3$ by modifying its crystal structure via zerovalent metal intercalation without significant PhP degradation
. 
Crystals of $\alpha$-MoO$_3$ intercalated with tin, copper, and silver were analyzed by photoinduced force microscopy (PiFM),\cite{shcherbakov_photo-induced_2025} revealing the influence of 
metal atoms on the PhP dispersion and losses. Although some intercalants, such as copper, did not show significant effects on PhP behavior
, others demonstrated prominent changes. For example, 
tin 
causes the shift in PhP dispersion towards shorter wavevectors 
of up to $38.5\pm0.5\%$ at $\omega=910$~cm$^{-1}$ excitation frequency
. At the same time, silver intercalation 
enables spectral tuning of PhP by shrinking its RB. 
The intercalated $\alpha$-MoO$_3$ exhibits low excess losses, with the average lifetimes of $\tau_{\rm Sn}=1.9 \pm0.8$~ps and $\tau_{\rm Ag}=1.4 \pm 0.3$~ps for tin and silver intercalation, respectively, similar to that of pristine $\alpha$-MoO$_3$ ($\tau_0=1.6 \pm 0.2$~ps). 
An analytical model of PhP dispersion suggests that the change in PhP behavior is caused by an increase in dielectric permittivity after intercalation, in agreement with the density functional theory results. Our findings 
provide an attractive pathway to reconfigurable mid-infrared nanophotonic devices for programmable planar optics.


\subsection{Intercalation tuning of PhPs dispersion}
 
The $\alpha$-MoO$_3$ samples 
were synthesized by a two-step process using the hydrothermal method followed by water vapor transport \cite{Reed2019}. Zerovalent metal intercalation in the van der Waals gap of $\alpha$-MoO$_3$ was conducted via a disproportionation redox reaction to generate zerovalent atoms by \citeauthor{Koski2012} \cite{Reed2019,Reed2024,Reed2024+} Intercalation concentrations, detected with energy-dispersive X-ray spectroscopy (EDX), were kept low at $\leq$1 atm \% to preserve the overall crystal symmetry; see Methods for more details on synthesis and intercalation of $\alpha$-MoO$_3$. 
 
The polaritonic response in pristine and intercalated $\alpha$-MoO$_3$ was studied using 
PiFM
, since PhPs are high-momentum light–matter waves that cannot be excited and characterized by free-space excitation. PiFM provides direct mapping of nanoscale near-field features and sample topography at the same time by illuminating the tip and sample with modulated mid-infrared light (Fig. \ref{fig1}a). 
PhP were characterized within the spectral range of the $\omega = 820$~cm$^{-1}$ – 970 cm$^{-1}$ RB of $\alpha$-MoO$_3$, where PhPs are known to propagate along the [100] crystallographic direction \cite{alvarez-perez_infrared_2020}. Figure \ref{fig1}b shows the topography and PiFM images at the $\omega=885$~cm$^{-1}$ excitation frequency of a 120-nm-thick $\alpha$-MoO$_3$, revealing interference fringes caused by the tip-launched polaritons reflected from the flake edges.

Figure~\ref{fig1}c shows a PiFM image of a 120-nm-thick 1 atm\% tin-intercalated $\alpha$-MoO$_3$ (Sn-MoO$_3$), demonstrating changes in the periodicity of the PhP interference pattern compared to pristine $\alpha$-MoO$_3$. The thickness of pristine and intercalated $\alpha$-MoO$_3$ was kept the same, as it is known that PhP dispersion depends on the flake thickness\cite{ma_-plane_2018}. 
Figure \ref{fig1}d shows the PiFM intensity line profiles, which were fitted to the exponentially decaying sinusoidal function \cite{ma_-plane_2018} :

\begin{equation}
    y=y_0+Ae^{-x/L}\text{sin}[\pi(x-x_c)/w],
    \label{eq1}
\end{equation}

where the PhP wavelength $\lambda_{\mathrm {PhP}}$ was extracted 
to be $\lambda_{\mathrm {PhP}} = 2w=1.21~\mu$m in pristine $\alpha$-MoO$_3$ compared to $\lambda_{\mathrm {PhP}}=1.47~\mu$m in Sn-MoO$_3$ at $\omega=885$~cm$^{-1}$
.


PiFM signal values are not directly comparable across measurements because the it is sensitive to experimental parameters such as tip and laser alignment.
For a quantitative analysis of the PhP properties, we compared the following figures of merit: group velocity $v_g$, propagation length $L$, and lifetime $\tau$. 
Figure \ref{fig1}e shows the dispersion $\omega(k_{\mathrm {PhP}})$, where $k_{\mathrm {PhP}} = 2\pi/\lambda_{\mathrm {PhP}}$ is the PhP wavevector along the [100] direction, extracted from PiFM images by sweeping the incident frequency, and fitted to $y=ax^b$. The fitting coefficients 
are $a_0=920.6\pm0.3$ and $b_0=0.058\pm0.001$ for a pristine, and $a_{\mathrm {Sn}}=942.0\pm3.5$ and $b_{\mathrm {Sn}}=0.068\pm0.004$ for a tin-intercalated samples, respectively. The relative shift in PhP dispersion $\Delta k/k = 21.5\pm4.5$\% towards lower PhP wavevectors was observed at $\omega=860$~cm$^{-1}$ after tin intercalation, where $\Delta k=k_{\text{$\alpha$-MoO$_3$}}-k_{\text{Sn-MoO$_3$}}$. At  $\omega=910$~cm$^{-1}$, $\Delta k/k = 38.5 \pm 0.5$\%
.
The PhP group velocity $v_g$ = d$\omega$/d$\text{Re}(k)$ in pristine and intercalated $\alpha$-MoO$_3$ is plotted in Fig. \ref{fig1}f. Here, $v_{g,\mathrm{Sn}} = 10^{-2}c$ for Sn-MoO$_3$ and $v_{g,0} = 7 \times 10^{-3}c$ for $\alpha$-MoO$_3$ at $\omega=880$~cm$^{-1}$, where $c$ is the speed of light in a vacuum. The PhP lifetime can be derived by $\tau=L/v_g$, where the propagation length $L$ is extracted from Eq.(\ref{eq1}), resulting in $L_0=1.2\pm0.2~\mu$m for pristine $\alpha$-MoO$_3$ and $L_{\mathrm{Sn}}=2.6\pm1.3~\mu$m for Sn-MoO$_3$ at $\omega=880$~cm$^{-1}$. The calculated average lifetime of PhP in pristine $\alpha$-MoO$_3$ $\tau_0=1.6 \pm 0.2$~ps is comparable to that in Sn-MoO$_3$ $\tau_{\mathrm{Sn}}=1.9 \pm 0.8$~ps, indicating minimum degradation of PhP after the intercalation process. 
The results shown above demonstrate that tin intercalation of $\alpha$-MoO$_3$ enables an effective and non-destructive approach to manipulate PhP dispersion, surpassing other PhP modulation approaches (see Table S1 for comparison details).

\begin{figure}[hbt!]
  \includegraphics[height=7.5cm]{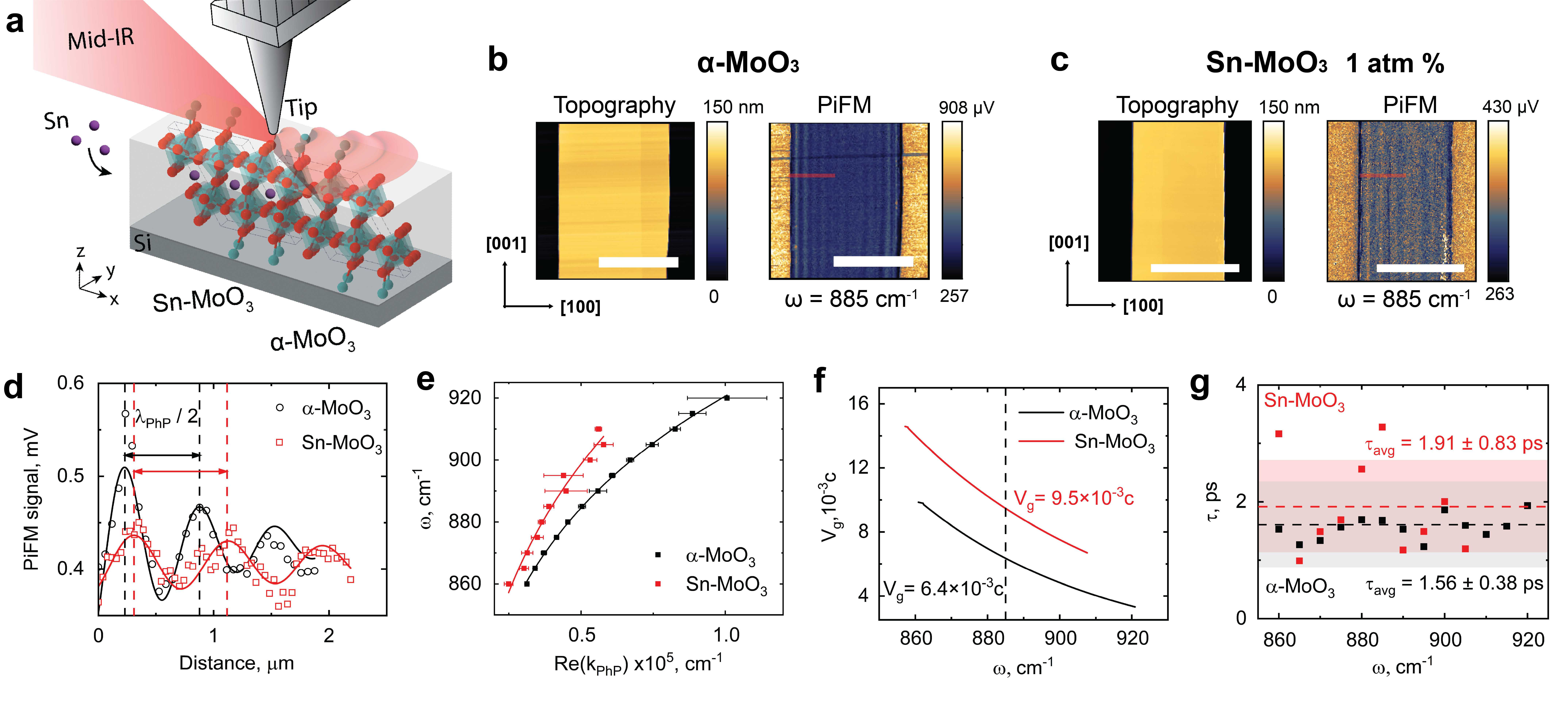}
  \caption{Intercalation tuning of phonon polaritons (PhPs) in $\alpha$-MoO$_3$. \textbf{a}, Schematic of tin intercalation of $\alpha$-MoO$_3$ and PiFM measurements of PhP propagation. \textbf{b}, Topography and PiFM images of a 120-nm-thick pristine $\alpha$-MoO$_3$ at $\omega$=885 cm$^{-1}$ excitation. \textbf{c}, Topography and PiFM images of a 120-nm-thick Sn-MoO$_3$ at $\omega$=885 cm$^{-1}$ excitation. Scale bars are 5 $\mu$m. \textbf{d}, PiFM intensity profiles extracted from \textbf{b} and \textbf{c}. $\lambda_{\mathrm {PhP}}$ is the PhP wavelength. \textbf{e}, Shift in PhP dispersion after tin intercalation of $\alpha$-MoO$_3$. \textbf{f}, PhP group velocities extracted from the fitting results in \textbf{e}. \textbf{g}, Comparison of the PhPs lifetime in pristine (black) and Sn-MoO$_3$ (red).} 
  \label{fig1}
\end{figure}

\begin{figure}
  \includegraphics[height=7.5cm]{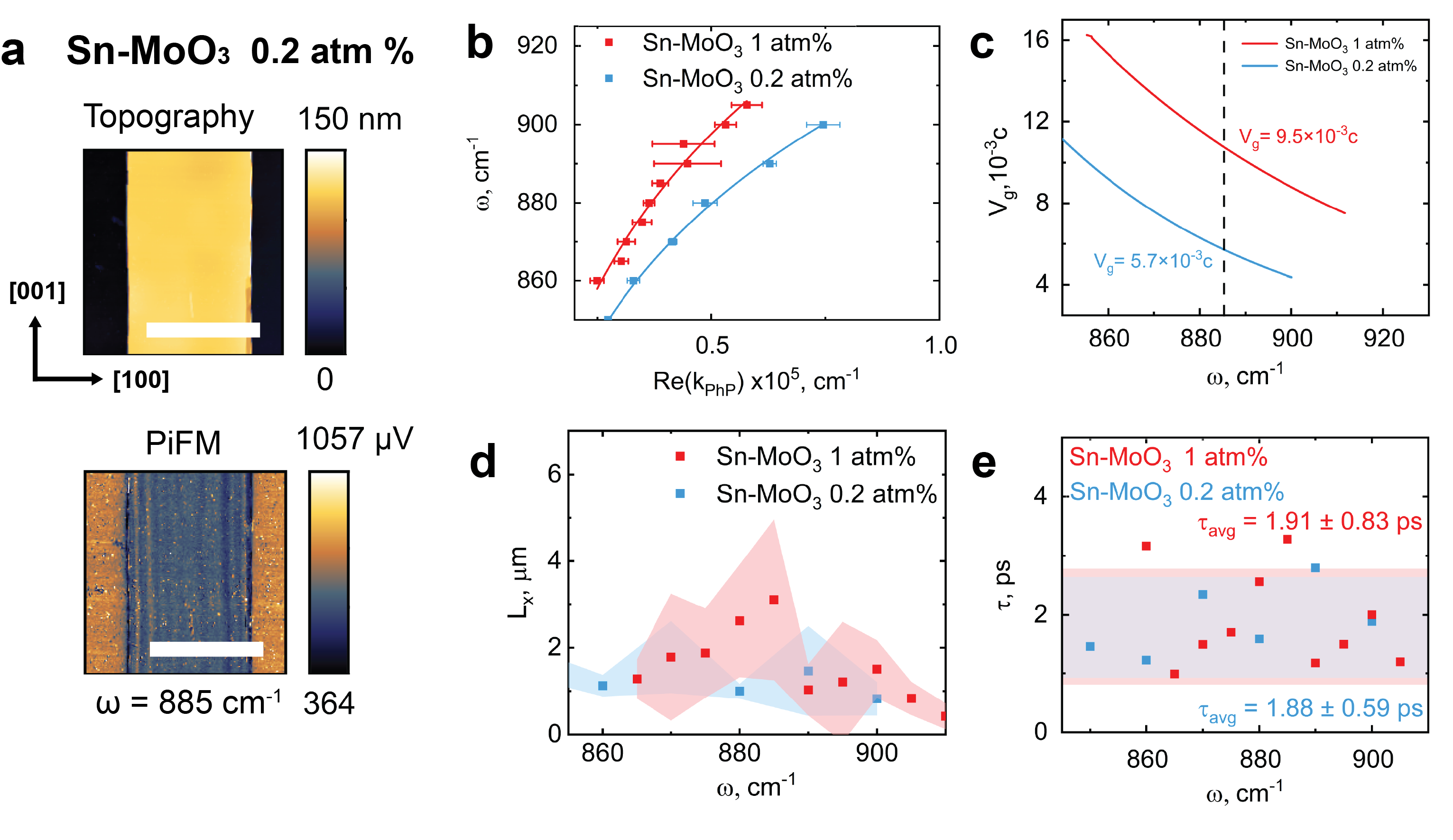}
  \caption{Concentration-dependent tuning of PhP in Sn-MoO$_3$. \textbf{a}, Topography and PiFM images at 885 cm$^{-1}$ excitation of 115 nm-thick 0.2 atm\% Sn-MoO$_3$. Scale bars are 5 $\mu$m. \textbf{b-e}, Comparison of PhP dispersion (\textbf{b}), group velocity (\textbf{c}), propagation length (\textbf{d}) and lifetime (\textbf{e}) of 120-nm-thick 1 atm\% Sn-MoO$_3$ (red) and 115-nm-thick 0.2 atm\% Sn-MoO$_3$ (blue).}
  \label{fig2}
\end{figure}

To explore the effect of intercalant concentration on PhP tunability, Sn-MoO$_3$ samples with a lower tin concentration of 0.2 atm\% were prepared by decreasing the time of the intercalation process. EDX spectra demonstrated a decrease in the tin signal (fig. S1), confirming that intercalation concentration decreases with a shorter intercalation time.
Figure \ref{fig2}a shows the topography and PiFM image of 115-nm-thick 0.2 atm\% Sn-MoO$_3$ at $\omega=885$~cm$^{-1}$. By comparing the PhP dispersion of 0.2~atm\% Sn-MoO$_3$ to 1~atm\% Sn-MoO$_3$, a gradual shift towards a higher PhP wavevector was found when the intercalant concentration was decreased (Fig. \ref{fig2}b). The decrease in the concentration of the intercalant brings the dispersion curve of PhP closer to one of the pristine $\alpha$-MoO$_3$ (fig. S2), diminishing the effect of the PhP modulation.
Figures \ref{fig2}c,d, and e compare the PhP group velocity, propagation length, and lifetime for two different concentrations of tin. The PhP group velocity decreases with decreasing concentration, while the average lifetime remains the same 
at $1.9 \pm 0.6$~ps against 
$1.9 \pm 0.8$~ps in 0.2 atm\% and 1 atm\% Sn-MoO$_3$, respectively.

\subsection{Intercalation species effect on PhP spectral tuning}

\begin{figure}[hbt!]
  \includegraphics[height=8.5cm]{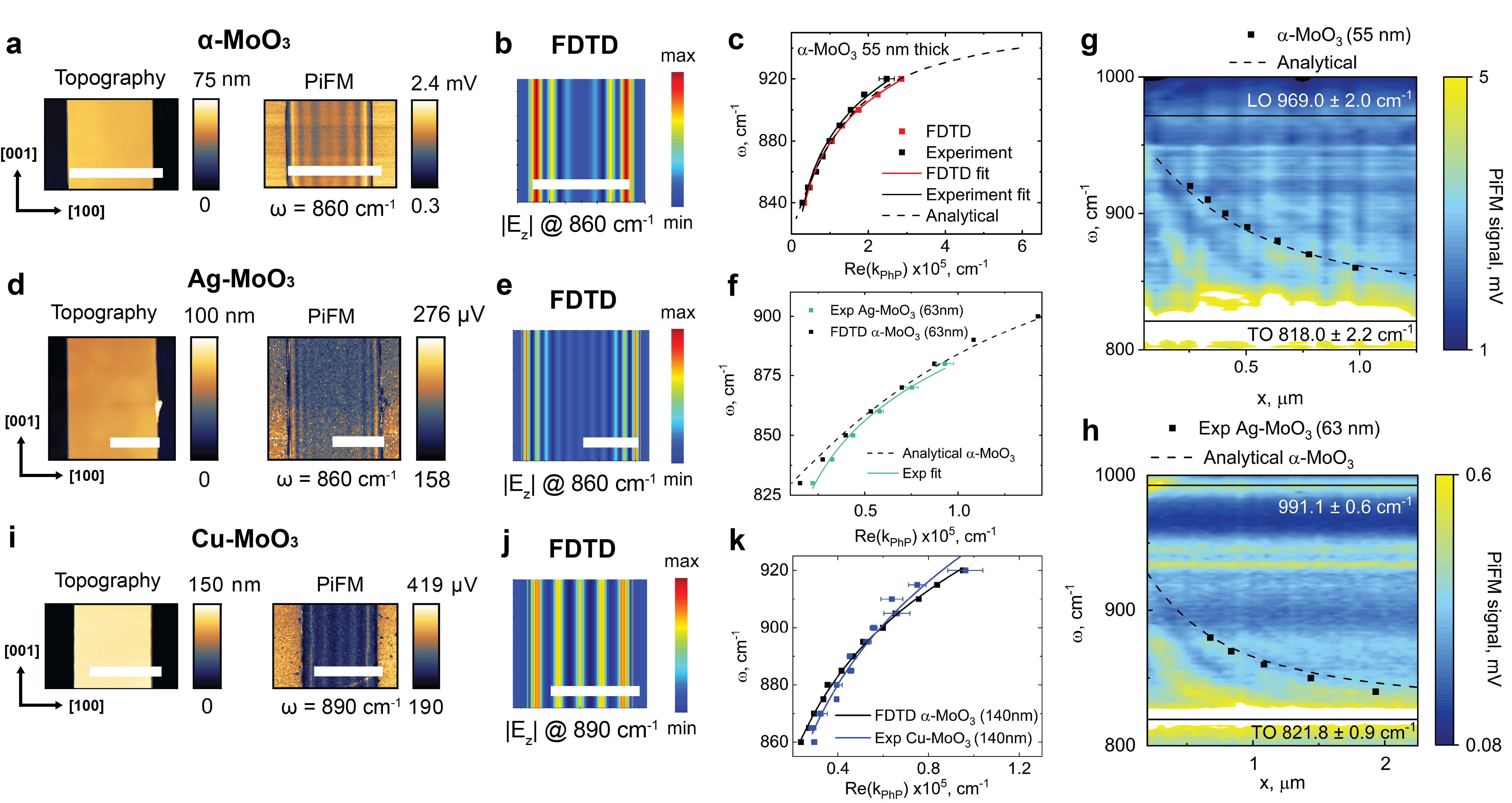}
  \caption{Analysis of intercalation species effect on PhP. \textbf{a}, Topography and PiFM images at 860 cm$^{-1}$ excitation of 55 nm-thick $\alpha$-MoO$_3$. \textbf{b}, FDTD simulated out-of-plane electric field distribution at 860 cm$^{-1}$ frequency of 55-nm-thick pristine $\alpha$-MoO$_3$. \textbf{c}, Experimental, analytical and simulated PhP dispersion of $\alpha$-MoO$_3$. \textbf{d}, Topography and PiFM images at 860 cm$^{-1}$ excitation of 63-nm-thick Ag-MoO$_3$. \textbf{e}, FDTD simulated out-of-plane electric field distribution at 860 cm$^{-1}$ frequency of 63-nm-thick pristine $\alpha$-MoO$_3$. \textbf{f}, Comparison of PhP dispersion of Ag-MoO$_3$ and simulated pristine $\alpha$-MoO$_3$. \textbf{g}, Spectral map along [100] 
  of $\alpha$-MoO$_3$. \textbf{h}, Spectral map along [100] 
  of Ag-MoO$_3$. \textbf{i}, Topography and PiFM images at 890 cm$^{-1}$ excitation of 140-nm-thick Cu-MoO$_3$. \textbf{j}, FDTD simulated out-of-plane electric field distribution at 890 cm$^{-1}$ frequency of 140-nm-thick pristine $\alpha$-MoO$_3$. \textbf{k}, Comparison of PhP dispersion of Cu-MoO$_3$ and simulated pristine $\alpha$-MoO$_3$. Scale bars are 3 $\mu$m.}
  \label{fig3}
\end{figure}

To investigate the PhP tuning possibilities by various metal-ion intercalations, we intercalated $\alpha$-MoO$_3$ with silver and copper. The concentration of intercalants was kept 
at 1 atm\% for both silver (Ag$_{0.04}$MoO$_3$) and 
copper (Cu$_{0.01}$MoO$_3$). 
The thicknesses of Ag-MoO$_3$ (63 nm) and Cu-MoO$_3$ (140 nm) are different from pristine $\alpha$-MoO$_3$ (120 nm) 
and, since the PhP dispersion depends on the flake thickness\cite{ma_-plane_2018}, finite-difference time-domain (FDTD) simulations using Ansys Lumerical (see Methods) 
were used to compare the PhP dispersion 
in Ag-MoO$_3$ and 
Cu-MoO$_3$ flakes to the 
pristine $\alpha$-MoO$_3$ of the same thickness. First, $\alpha$-MoO$_3$ was simulated to confirm the reliability of the simulated model. In simulation, the $\alpha$-MoO$_3$ flake with the same thickness as the experimental one (Fig. \ref{fig3}a) was placed on the silicon substrate and illuminated by a plane wave. A 2D monitor was placed on the surface of flake to measure the out-plane electric field distribution. Figure \ref{fig3}b shows the calculated out-of-plane electric field of 55 nm-thick pristine $\alpha$-MoO$_3$, which demonstrates characteristic interference fringes with the periodicity of the edge-launched PhP wavelength. The simulated PhP wavelength was exported from the out-plane monitor. 
The resulting PhP dispersion of $\alpha$-MoO$_3$ was consistent with the experimental results (Fig. \ref{fig3}c). 

Figures \ref{fig3}d and i demonstrate the topography and PiFM images for a Ag-MoO$_3$ sample at $\omega=860$~cm$^{-1}$ and a Cu-MoO$_3$ sample at $\omega=890$~cm$^{-1}$, respectively. The PhP dispersion of 63 nm-thick Ag-MoO$_3$ shows no observable shift compared to the dispersion of FDTD-simulated 63-nm-thick pristine $\alpha$-MoO$_3$ (Fig. \ref{fig3}f), exported from the out-of-plane monitor (Fig. \ref{fig3}e). However, 
the PhP response disappeared at excitation frequencies $\omega>880$~cm$^{-1}$ as seen from the absence of data points in the dispersion plot, which was not the case for the pristine $\alpha$-MoO$_3$ (fig. S3).

The 
difference between PhPs in Ag-MoO$_3$ and $\alpha$-MoO$_3$ can be seen from PiFM spectrally-resolved maps in Figure \ref{fig3}g, h. By moving the tip along [100] starting from the edge into the flake and recording PiFM spectra at each tip position, we observe fringe periodicity as a function of the excitation frequency; see Methods for more details on the measurement of the PiFM spectral map. First, the spectral map of a 55-nm-thick pristine $\alpha$-MoO$_3$ was measured using this approach (Fig. \ref{fig3}g). The spectra demonstrate a series of signal maxima with different spacings, indicating that the PhP wavelength changes at different excitation frequencies. Within the [100] RB, 
the PhP wavelength decreases with increasing the excitation frequency. The same trend is observed in the PhP dispersion measurement, 
well aligned with the experimental data points (black squares) and analytical curve (dashed lines) in Fig.~\ref{fig3}g. The series of signal maxima are present within the [100] RB limits \cite{taboada-gutierrez_broad_2020} corresponding to the LO-TO phonon frequencies at $970-820$~cm$^{-1}$, respectively \cite{Taboada2024,wu_chemical_2020}. In stark contrast, the signal maxima of the Ag-MoO$_3$ spectral map vanish around 890 cm$^{-1}$ 
(Fig. \ref{fig3}h), confirming the absence of PhP signal and suggesting potential modifications to the LO-TO structure of the sample. 
Finally, the PhP dispersion of Cu-MoO$_3$ with the same intercalant concentration as tin and silver has not demonstrated any difference compared to pristine $\alpha$-MoO$_3$ (Fig. \ref{fig3}i-k, fig. S4). Therefore, the choice of intercalant material provides broad spectral control and switching of the PhP.

\subsection{Density Functional Theory and Analytical Calculation Results}

\begin{figure}
  \includegraphics[height=15cm]{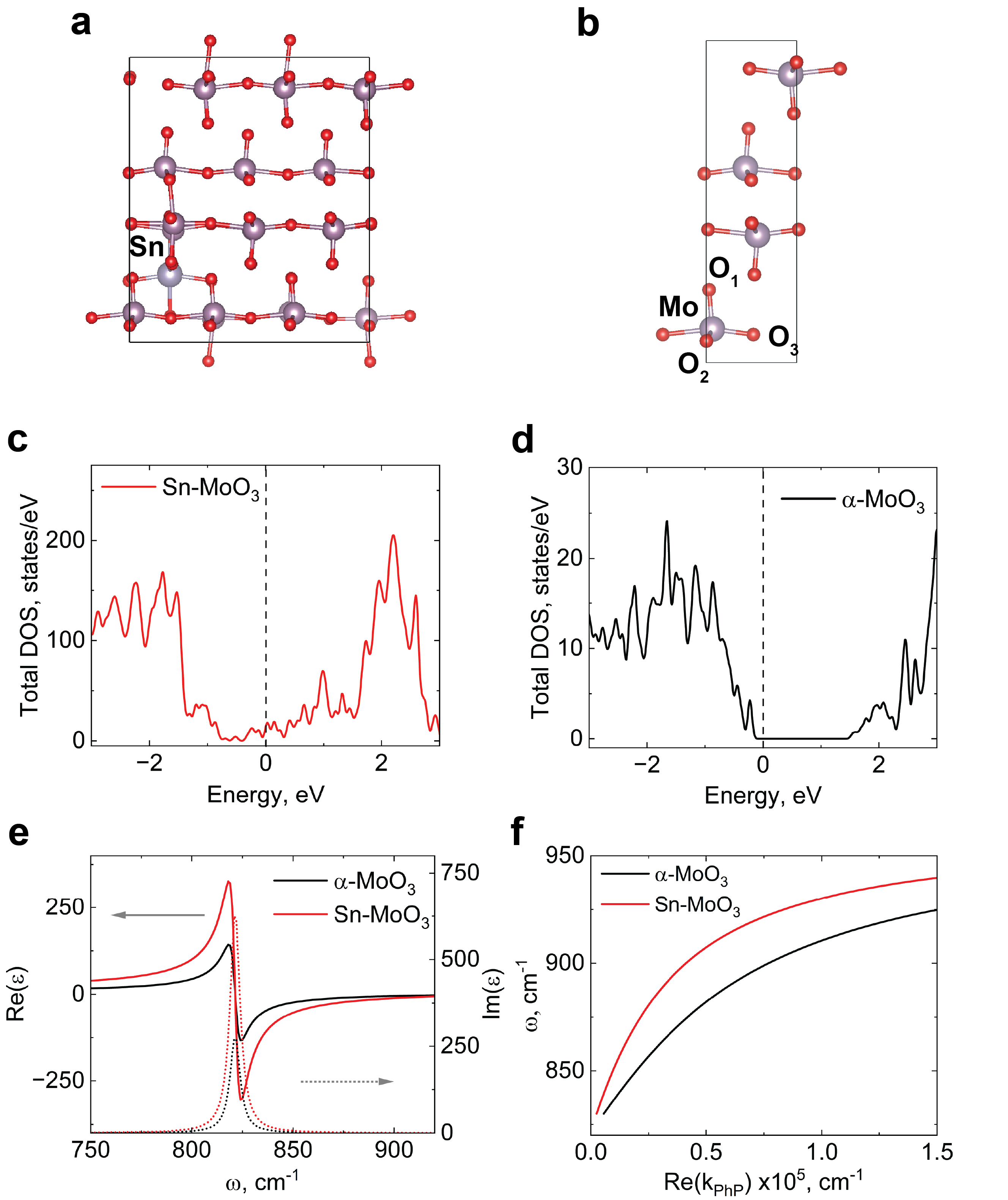}
  \caption{DFT and analytical dispersion of PhP. \textbf{a}, Relaxed crystal structure of Sn-MoO$_3$, with one intercalant placed in a 3 $\times$ 3 $\times$ 1 supercell. \textbf{b},  Primitive cell of $\alpha$-MoO$_3$ used in pristine DFT calculations. \textbf{c}, Total DOS of Sn-MoO$_3$. \textbf{d}, Total DOS of $\alpha$-MoO$_3$. \textbf{e}, Dielectric permittivity $\varepsilon_x$ model of $\alpha$-MoO$_3$ and Sn-MoO$_3$ based on DFT calculated $\varepsilon^x_\infty$. \textbf{f},  Analytical PhP dispersion of $\alpha$-MoO$_3$ with original $\hat{\varepsilon}_\infty$ compared to Sn-MoO$_3$ with DFT calculated $\hat{\varepsilon}_\infty$.}
  \label{fig4}
\end{figure}

To understand the reason behind the observed changes in the PhP behavior during intercalation, density functional theory (DFT) was used to calculate the electronic structure and static dielectric permittivity of $\alpha$-MoO$_3$, along with an tin-intercalated structure. 
The DFT model consists of a 3x3x1 supercell with one tin atom placed in the vdW gap of $\alpha$-MoO$_3$ (Fig. \ref{fig4}a,b) at the most energetically stable doping site \cite{xu_electronic_2020}. The DFT-relaxed geometry of Sn-MoO$_3$ shows 0.9\% lattice expansion along [100], 2.2\% contraction along [001] and 7.4\% expansion along [010], which is consistent with previous reports \cite{huang_impact_2014, xu_electronic_2020}. Also, the Sn-MoO$_3$ shows increases in static dielectric constants compared to the pristine structure, from $\varepsilon^x_\infty=6.7$ to 13.3, $\varepsilon^y_\infty=6.5$ to 13.2, and from $\varepsilon^z_\infty=4.5$ to 4.8, as obtained from the density functional perturbation theory. This finding is consistent with the calculated density of states (DOS) for Sn-MoO$_3$ (Fig. \ref{fig4}c) and $\alpha$-MoO$_3$ (Fig. \ref{fig4}d), which reveals 
intermediate states in the band gap. 
The new electronic states near the Fermi level boost polarizability due to the polarization of induced dipoles \cite{rajapakse_intercalation_2021,doi:10.1021/acsami.0c02301} and, in turn, increase 
$\varepsilon_\infty$\cite{huang_impact_2014, xu_electronic_2020}. Although metal intercalation 
introduces free carriers, we expect any plasmon-polariton contribution to be negligible given the low intercalant concentration.

To quantitatively estimate the shift of PhP dispersion induced by the modulation of static dielectric permittivity calculated by DFT, an analytical model of PhP dispersion\cite{alvarez-perez_infrared_2020} is considered (see details in note S1). Following this model, we plot the dielectric permittivity and analytical PhP dispersion for pristine $\alpha$-MoO$_3$ and Sn-MoO$_3$. Figure \ref{fig4}e shows the comparison between dielectric permittivity $\varepsilon_x(\omega)$ of pristine $\alpha$-MoO$_3$ and Sn-MoO$_3$ with DFT-calculated static dielectric constant by adjusting this parameter in the dielectric permittivity model. This permittivity shift alters the PhP dispersion, lengthening the PhP wavelength and shifting the dispersion curve to lower in-plane wavevectors, as shown by analytical modeling (Figure \ref{fig4}f). For the analytical PhP dispersion of Sn-MoO$_3$, the DFT calculated static dielectric tensor $\hat{\varepsilon}_\infty$ was used, while other parameters such as $\omega_{TO}$ and $\omega_{LO}$ were kept the same after tin intercalation according to the obtained spectral maps (fig. S6), as well as PhP lifetime (Fig. \ref{fig1}g) which 
suggests little variation in the damping factor $\gamma$. 
 For Sn-MoO$_3$ (Fig. \ref{fig4}f, red curve), the average shift of $63\pm5$\% towards the lower wavevector values was found relative to pristine $\alpha$-MoO$_3$ (Fig. \ref{fig4}f, black curve). The estimated shift in the analytical PhP dispersion with the DFT-predicted permittivity appears to be larger than the experimentally observed one due to the tripled intercalant concentration in the calculation (Sn$_{0.03}$MoO$_3$) compared to the experimental sample (Sn$_{0.01}$MoO$_3$). A DFT calculation of a 5x5x1 supercell would be required to properly simulate 
the experimental sample; however, significantly higher structure complexity and longer computational time would be required. Under the conjecture that the permittivity shift changes linearly with the intercalant concentration, the adjusted calculated PhP dispersion shift is $21\pm5\%$, which is comparable to the experimentally observed average shift of $23.4\pm7.2\%$.

The Lorentz oscillator model of dielectric permittivity in Eq.(S3) 
suggests two primary mechanisms for the PhP tunability, assuming the damping factor variations are insignificant. 
First, the static dielectric constant can be changed by intercalation 
via 
controlling the band structure and Fermi level \cite{huang_impact_2014, rajapakse_intercalation_2021, peelaers_controlling_2017}, likewise in case of Sn-MoO$_3$.  Second, the magnitude of the LO-TO splitting, which reflects the oscillator strength of the given phonon mode \cite{Galiffi2023}, can be tuned by changing the intercalant species, because different ionic radii 
lead to different charge distributions and the electric field introduced by the dopant charge, causing changes to the local lattice 
\cite{xu_electronic_2020}. Silver intercalation could potentially change PhP behavior caused by the second route.

Experimental elemental quantification of the intercalated 
material 
challenging due to the small intercalant concentrations. 
Instead of EDX, electron probe microanalysis (EPMA) with wavelength-dispersive spectrometers (WDS) 
may be used for its 
sensitivity 
(10-100 
ppm 
against EDX with a detection limit of 
$\sim 1000-10,000$ ppm) and better accuracy of $\pm1-2$\% for EPMA/WDS compared to $\pm2-10$\% for EDX \cite{https://doi.org/10.1002/cjce.25712,10.1017/S1431927621000192}. In addition, 
a real-time, 
in situ method of monitoring the structural changes during the intercalation process, such as time-resolved XRD \cite{doi:10.1021/acs.chemrev.3c00331}, would be helpful in controllable PhP modification. 


As our data suggests, intercalation by zerovalent metals 
manipulates PhPs in $\alpha$-MoO$_3$. 
Chemically, the metal intercalant most likely interacts with the double bonded oxygen in $\alpha$-MoO$_3$. The nature of that interaction gives insight into possible choice of intercalant. Therefore, the intercalation tuning of PhP can be envisioned in other oxides
, such as $\alpha$-V$_2$O$_5$ \cite{taboada-gutierrez_broad_2020}, YVO$_4$ \cite{liu_long-range_2025}, CdWO$_4$ \cite{hu_real-space_2023}, $\beta$-Ga$_2$O$_3$ and $\alpha$-SiO$_2$ \cite{passler_hyperbolic_2022}.
In addition, it is possible to introduce more than one type of intercalant in the same sample, which can give a more controllable way to manipulate PhP. For example, the synergy of completely different charge distributions around tin and cobalt in co-intercalated $\alpha$-MoO$_3$ leads to extended gap states across the forbidden band \cite{xu_electronic_2020}, which potentially can enhance electronic conductivity, even stronger than the single-atomic doping system.
Further studies of intercalation reaction stage, intercalant species, host materials, and real-time in situ characterization \cite{rajapakse_intercalation_2021} will allow deeper understanding of how structural modifications affect PhP behavior.

\subsection{Conclusion}

We have 
demonstrated zerovalent metal intercalation as an efficient chemical approach to manipulate PhPs in a vdW crystal. 
Near-field imaging of Sn-MoO$_3$ revealed a relative shift in PhPs dispersion of up to $38.5\pm0.5$\%, 
while PiFM spectral mapping of Ag-MoO$_3$ showed 
PhP disappearance at excitation frequencies above 890 cm$^{-1}$ compared to 920 cm$^{-1}$ for pristine $\alpha$-MoO$_3$. The lifetime of tin- and silver-intercalated samples (1.9±0.8 ps and 1.4±0.3 ps, respectively) remained quantitatively comparable to pristine $\alpha$-MoO$_3$ (1.6±0.2 ps), indicating 
minimum PhPs degradation after the intercalation
. The shift in PhP dispersion 
can be explained by a modulation of the static dielectric permittivity, as 
supported by a 
PhP analytical model and 
DFT-calculated permittivity doubling 
along [100] in Sn-MoO$_3$
. Further tuning of PhP could be achieved through a variety of metal atoms as well as co-doped intercalation in different host layered materials and could enable PhPs applications in developing and designing programmable nanophotonic devices and scalable planar optics.

\section{Methods}

\subsection{Sample preparation}


$\alpha$-MoO$_3$ flakes are synthesized using the hydrothermal method of \citeauthor{Li2006} \cite{Li2006} $\alpha$-MoO$_3$ flakes were grown by a two step process. First, precursor $\alpha$-MoO$_3$ nanoribbons were synthesized hydrothermally \cite{Wang2015}. These nanoribbons were used as the source for growth of larger crystals. 
Laterally larger $\alpha$-MoO$_3$ 2D layered crystals 
were grown from the precursor using physical vapor deposition with water vapor as a transport agent. Most of the water supernatant was removed, and the 
$\alpha$-MoO$_3$ flakes 
were semi-dried under nitrogen gas for 1 h. $\alpha$-MoO$_3$ flakes, with some remaining water, were placed in a 10 mm diameter open quartz boat in the center of a 25 mm quartz tube in a single-zone tube furnace. Both ends of the tube were sealed. The furnace was heated to 850°C at its center. $\alpha$-MoO$_3$ was vaporized and transported downstream 14.3 cm from the center to both sides of the tube. The flakes formed a spider-web structure vertically spanning the quartz tube and were collected with tweezers.
The $\alpha$-MoO$_3$ flakes were mechanically transferred to a silicon substrate by exfoliation using Nitto blue tape.


Tin, silver, and copper were intercalated into the van der Waals gap of $\alpha$-MoO$_3$ using the methods of \citeauthor{Koski2012} \cite{Reed2019,Reed2024,Reed2024+} Intercalation of tin proceeds via the disproportionation redox reaction of Sn(II). The $\alpha$-MoO$_3$ flakes were placed unconstrained in a 10 mM solution of SnCl$_2$ and 100 mM tartaric acid in acetone just under reflux for 1 h \cite{Koski2012}. Copper was intercalated via the disproportionation redox reaction of Cu(I). Flakes were placed into a solution containing 5 mL of acetone and 0.13 g of tetrakis(acetonitrile)copper(I) hexafluorophosphate; the solution was kept under reflux for 4 h. Glassware for copper intercalation reactions were prepared before the reaction by placing in distilled water (pH around 5.5-6.5) overnight followed by drying for 24 hours in an oven at 100°C. Zero-valent cobalt was intercalated by the carbonyl decomposition of dicobalt octacarbonyl.
Intercalated flakes were rinsed repeatedly with acetone and ethanol following intercalation by suspending the flakes in acetone and then filtering them from solution. 
The intercalated atomic percentage of metal was kept on the order of around 0.2 atm\% and 1 atm\% tin (Sn$_{0.01}$MoO$_3$), 1 atm\% silver (Ag$_{0.04}$MoO$_3$), and 1 atm\% copper (Cu$_{0.01}$MoO$_3$). Intercalated $\alpha$-MoO$_3$ flakes were mechanically transferred to a silicon substrate by exfoliation using Nitto blue tape. 


\subsection{Near-field optical imaging}

To image the PhP in real space, a photo-induced force microscope (PiFM) was used. PiFM detects localized near-field features of the investigated material as forced oscillations of cantilever, illuminated with mid-infrared light. The near-field optical distributions of the sample can be mapped simultaneously with its topography. PiFM measurements were performed using a commercial Molecular Vista (Vista One) setup equipped with a pulsed mid-infrared quantum cascade laser (Block Engineering), whose emission frequency can be tuned between 782 and 1920 cm$^{-1}$ with 1 cm$^{-1}$ spectral resolution. The p-polarized light from the laser was focused on a Pt/Ir-coated tip, with a resonant frequency of around 300 kHz, operating at 5\% power in tapping mode. The laser was modulated at the frequency difference between the first and second flexural eigenmodes of the cantilever. The second mechanical eigenmode was used in the AFM feedback loop to control the tip-sample distance, with typical tapping modulation amplitude of 1 nm and typical set point of 75\%. The amplitude of the light-induced oscillations of the first cantilever eigenmode was recorded as a PiFM signal.
The spectral map was measured by moving tip along the line parallel to the [100] crystallographic direction of $\alpha$-MoO$_3$ from the edge into the flake. The PiFM spectra have been recorded at each tip position with 50 nm step 
at 5\% laser power. 


\subsection{Finite-Difference Time-Domain Simulations}
We conducted full-wave electromagnetic simulations using a finite-difference time-domain (FDTD) method on commercially available Ansys Lumerical FDTD software. $\alpha$-MoO$_3$ flakes with a geometry that matches the experimental ones on Si substrate were simulated. Plane wave source polarized along [100] direction was used in simulations and flake edges act as PhP launchers. We placed a 2D field profile monitor on top of the flake surface to measure E$_z$. The dispersion was then calculated from extracted E$_z$ line profiles. 

\subsection{Density Functional Theory}

DFT calculations were performed using the Vienna Ab initio Simulation Package (VASP) \cite{kresse1996efficient,kresse1993ab} with projector augmented wave (PAW) pseudopotentials for Mo (4s2 4p6 5s1 4d5) and O (2s2 2p4). The van der Waals interactions were treated with the DFT-D3 correction \cite{grimme2011effect}. A Hubbard U term of 5 eV was applied to Mo to better account for the localized d-electrons. All geometries were relaxed until the force on each atom was less than 0.01 eV for stoichiometric cells and 0.05 eV for defect cells. A $\Gamma$-centered 12x12x3 k-mesh\cite{monkhorst1976special,langreth2005van,tong2021first} is used for the primitive cell calculations and 4x4x3 for the 3x3x1 supercells used in intercalated structures. In all cases, an energy cutoff of 600 eV is used, and electronic convergence is set to $10^{-5}$ for geometry relaxation and $10^{-6}$ for self-consistency calculations. Static dielectrics were obtained using density functional perturbation theory within VASP\cite{langreth2005van,tong2021first}.

\begin{acknowledgement}

\textbf{Funding:} M.S., M.A.S., and M.R.S. acknowledge support from the NSF (grant no. ECCS-2339271). J.B., W.H., and M.R.S. acknowledge support from DARPA (grant no. D22AP00153). K.Koski acknowledges support from the NSF (grant no. DMR-2202472)

\textbf{Author contributions:} M.S. and M.R.S. conceived the idea and designed the experiments.  K.K., M.S. and M.N. performed sample preparation and PiFM measurements. W.H. and R.W. performed DFT calculations. M.A.S. and M.S. performed analytical and FDTD calculations. K.K. performed sample synthesis and SEM and EDS measurements. J.B. and M.S. performed data analysis. M.S. and M.R.S. wrote the manuscript with contribution from all authors. 
M.R.S. supervised the project.  

\end{acknowledgement}

\nocite{zhao2022ultralow,taboada2024unveiling,zhou2023gate}




\begin{mcitethebibliography}{61}
\providecommand*\natexlab[1]{#1}
\providecommand*\mciteSetBstSublistMode[1]{}
\providecommand*\mciteSetBstMaxWidthForm[2]{}
\providecommand*\mciteBstWouldAddEndPuncttrue
  {\def\EndOfBibitem{\unskip.}}
\providecommand*\mciteBstWouldAddEndPunctfalse
  {\let\EndOfBibitem\relax}
\providecommand*\mciteSetBstMidEndSepPunct[3]{}
\providecommand*\mciteSetBstSublistLabelBeginEnd[3]{}
\providecommand*\EndOfBibitem{}
\mciteSetBstSublistMode{f}
\mciteSetBstMaxWidthForm{subitem}{(\alph{mcitesubitemcount})}
\mciteSetBstSublistLabelBeginEnd
  {\mcitemaxwidthsubitemform\space}
  {\relax}
  {\relax}

\bibitem[Low \latin{et~al.}(2016)Low, Chaves, Caldwell, Kumar, Fang, Avouris,
  Heinz, Guinea, Martin-Moreno, and Koppens]{Low2016}
Low,~T.; Chaves,~A.; Caldwell,~J.~D.; Kumar,~A.; Fang,~N.~X.; Avouris,~P.;
  Heinz,~T.~F.; Guinea,~F.; Martin-Moreno,~L.; Koppens,~F. Polaritons in
  layered two-dimensional materials. \emph{Nature Materials} \textbf{2016},
  \emph{16}, 182--194\relax
\mciteBstWouldAddEndPuncttrue
\mciteSetBstMidEndSepPunct{\mcitedefaultmidpunct}
{\mcitedefaultendpunct}{\mcitedefaultseppunct}\relax
\EndOfBibitem
\bibitem[Taboada-Gutiérrez \latin{et~al.}(2024)Taboada-Gutiérrez, Zhou,
  Tresguerres-Mata, Lanza, Martínez-Suárez, Álvarez Pérez, Duan, Martín,
  Vélez, Prieto, Bercher, Teyssier, Errea, Nikitin, Martín-Sánchez,
  Kuzmenko, and Alonso-González]{Taboada2024}
Taboada-Gutiérrez,~J. \latin{et~al.}  Unveiling the Mechanism of
  Phonon-Polariton Damping in $\alpha$-MoO3. \emph{ACS Photonics} \textbf{2024},
  \emph{11}, 3570--3577\relax
\mciteBstWouldAddEndPuncttrue
\mciteSetBstMidEndSepPunct{\mcitedefaultmidpunct}
{\mcitedefaultendpunct}{\mcitedefaultseppunct}\relax
\EndOfBibitem
\bibitem[Wu \latin{et~al.}(2022)Wu, Duan, Ma, Ou, Li, Alonso-González,
  Caldwell, and Bao]{Wu2022}
Wu,~Y.; Duan,~J.; Ma,~W.; Ou,~Q.; Li,~P.; Alonso-González,~P.;
  Caldwell,~J.~D.; Bao,~Q. Manipulating polaritons at the extreme scale in van
  der Waals materials. \emph{Nature Reviews Physics} \textbf{2022}, \emph{4},
  578--594\relax
\mciteBstWouldAddEndPuncttrue
\mciteSetBstMidEndSepPunct{\mcitedefaultmidpunct}
{\mcitedefaultendpunct}{\mcitedefaultseppunct}\relax
\EndOfBibitem
\bibitem[Basov \latin{et~al.}(2016)Basov, Fogler, and Abajo]{Basov2016}
Basov,~D.~N.; Fogler,~M.~M.; Abajo,~F. J. G.~D. Polaritons in van der Waals
  materials. \emph{Science} \textbf{2016}, \emph{354}\relax
\mciteBstWouldAddEndPuncttrue
\mciteSetBstMidEndSepPunct{\mcitedefaultmidpunct}
{\mcitedefaultendpunct}{\mcitedefaultseppunct}\relax
\EndOfBibitem
\bibitem[Foteinopoulou \latin{et~al.}(2019)Foteinopoulou, Devarapu, Subramania,
  Krishna, and Wasserman]{Foteinopoulou2019}
Foteinopoulou,~S.; Devarapu,~G. C.~R.; Subramania,~G.~S.; Krishna,~S.;
  Wasserman,~D. Phonon-polaritonics: Enabling powerful capabilities for
  infrared photonics. \emph{Nanophotonics} \textbf{2019}, \emph{8},
  2129--2175\relax
\mciteBstWouldAddEndPuncttrue
\mciteSetBstMidEndSepPunct{\mcitedefaultmidpunct}
{\mcitedefaultendpunct}{\mcitedefaultseppunct}\relax
\EndOfBibitem
\bibitem[Galiffi \latin{et~al.}(2023)Galiffi, Carini, Ni, Álvarez Pérez,
  Yves, Renzi, Nolen, Wasserroth, Wolf, Alonso-Gonzalez, Paarmann, and
  Alù]{Galiffi2023}
Galiffi,~E.; Carini,~G.; Ni,~X.; Álvarez Pérez,~G.; Yves,~S.; Renzi,~E.~M.;
  Nolen,~R.; Wasserroth,~S.; Wolf,~M.; Alonso-Gonzalez,~P.; Paarmann,~A.;
  Alù,~A. Extreme light confinement and control in low-symmetry
  phonon-polaritonic crystals. \emph{Nature Reviews Materials} \textbf{2023},
  \emph{9}, 9--28\relax
\mciteBstWouldAddEndPuncttrue
\mciteSetBstMidEndSepPunct{\mcitedefaultmidpunct}
{\mcitedefaultendpunct}{\mcitedefaultseppunct}\relax
\EndOfBibitem
\bibitem[Hu \latin{et~al.}(2020)Hu, Shen, Qiu, Alù, Dai, Hu, Alù, w~Qiu,
  Shen, and Dai]{Hu2020}
Hu,~G.; Shen,~J.; Qiu,~C.-W.; Alù,~A.; Dai,~S.; Hu,~G.; Alù,~A.; w~Qiu,~C.;
  Shen,~J.; Dai,~S. Phonon Polaritons and Hyperbolic Response in van der Waals
  Materials. \emph{Advanced Optical Materials} \textbf{2020}, \emph{8},
  1901393\relax
\mciteBstWouldAddEndPuncttrue
\mciteSetBstMidEndSepPunct{\mcitedefaultmidpunct}
{\mcitedefaultendpunct}{\mcitedefaultseppunct}\relax
\EndOfBibitem
\bibitem[Ma \latin{et~al.}(2024)Ma, Zhong, Dai, and Ou]{Ma2024}
Ma,~Y.; Zhong,~G.; Dai,~Z.; Ou,~Q. In-plane hyperbolic phonon polaritons:
  materials, properties, and nanophotonic devices. \emph{npj Nanophotonics}
  \textbf{2024}, \emph{1}, 1--14\relax
\mciteBstWouldAddEndPuncttrue
\mciteSetBstMidEndSepPunct{\mcitedefaultmidpunct}
{\mcitedefaultendpunct}{\mcitedefaultseppunct}\relax
\EndOfBibitem
\bibitem[Duan \latin{et~al.}(2021)Duan, Álvarez Pérez, Tresguerres-Mata,
  Taboada-Gutiérrez, Voronin, Bylinkin, Chang, Xiao, Liu, Edgar, Martín,
  Volkov, Hillenbrand, Martín-Sánchez, Nikitin, and
  Alonso-González]{duan_planar_2021}
Duan,~J. \latin{et~al.}  Planar refraction and lensing of highly confined
  polaritons in anisotropic media. \emph{Nature Communications} \textbf{2021},
  \emph{12}, 4325\relax
\mciteBstWouldAddEndPuncttrue
\mciteSetBstMidEndSepPunct{\mcitedefaultmidpunct}
{\mcitedefaultendpunct}{\mcitedefaultseppunct}\relax
\EndOfBibitem
\bibitem[Zheng \latin{et~al.}(2022)Zheng, Jiang, Xu, Wang, Huang, Ke, Zhang,
  Chen, and Deng]{zheng_controlling_2022}
Zheng,~Z.; Jiang,~J.; Xu,~N.; Wang,~X.; Huang,~W.; Ke,~Y.; Zhang,~S.; Chen,~H.;
  Deng,~S. Controlling and Focusing In-Plane Hyperbolic Phonon Polaritons in
  $\alpha$-MoO$_3$ with a Curved Plasmonic Antenna. \emph{Advanced Materials}
  \textbf{2022}, \emph{34}, 2104164\relax
\mciteBstWouldAddEndPuncttrue
\mciteSetBstMidEndSepPunct{\mcitedefaultmidpunct}
{\mcitedefaultendpunct}{\mcitedefaultseppunct}\relax
\EndOfBibitem
\bibitem[Liang \latin{et~al.}(2024)Liang, Zhu, Zhao, Wang, Gong, Zhang, and
  Wang]{liang_manipulation_2024}
Liang,~J.; Zhu,~J.; Zhao,~Y.; Wang,~S.; Gong,~Y.; Zhang,~Y.; Wang,~G.~P.
  Manipulation of In-Plane Hyperbolic Phonon Polaritons for Configurable
  Focusing. \emph{{ACS} Photonics} \textbf{2024}, \emph{11}, 5031--5038\relax
\mciteBstWouldAddEndPuncttrue
\mciteSetBstMidEndSepPunct{\mcitedefaultmidpunct}
{\mcitedefaultendpunct}{\mcitedefaultseppunct}\relax
\EndOfBibitem
\bibitem[Li \latin{et~al.}(2015)Li, Lewin, Kretinin, Caldwell, Novoselov,
  Taniguchi, Watanabe, Gaussmann, and Taubner]{li_hyperbolic_2015}
Li,~P.; Lewin,~M.; Kretinin,~A.~V.; Caldwell,~J.~D.; Novoselov,~K.~S.;
  Taniguchi,~T.; Watanabe,~K.; Gaussmann,~F.; Taubner,~T. Hyperbolic
  phonon-polaritons in boron nitride for near-field optical imaging and
  focusing. \emph{Nature Communications} \textbf{2015}, \emph{6}, 7507\relax
\mciteBstWouldAddEndPuncttrue
\mciteSetBstMidEndSepPunct{\mcitedefaultmidpunct}
{\mcitedefaultendpunct}{\mcitedefaultseppunct}\relax
\EndOfBibitem
\bibitem[Duan \latin{et~al.}(2025)Duan, Martín-Luengo, Lanza, Partel, Voronin,
  Tresguerres-Mata, Álvarez Pérez, Nikitin, Martín-Sánchez, and
  Alonso-González]{duan_canalization-based_2025}
Duan,~J.; Martín-Luengo,~A.~T.; Lanza,~C.; Partel,~S.; Voronin,~K.;
  Tresguerres-Mata,~A. I.~F.; Álvarez Pérez,~G.; Nikitin,~A.~Y.;
  Martín-Sánchez,~J.; Alonso-González,~P. Canalization-based
  super-resolution imaging using an individual van der Waals thin layer.
  \emph{Science Advances} \textbf{2025}, \emph{11}, eads0569\relax
\mciteBstWouldAddEndPuncttrue
\mciteSetBstMidEndSepPunct{\mcitedefaultmidpunct}
{\mcitedefaultendpunct}{\mcitedefaultseppunct}\relax
\EndOfBibitem
\bibitem[Dai \latin{et~al.}(2015)Dai, Ma, Andersen, Mcleod, Fei, Liu, Wagner,
  Watanabe, Taniguchi, Thiemens, Keilmann, Jarillo-Herrero, Fogler, and
  Basov]{dai_subdiffractional_2015}
Dai,~S.; Ma,~Q.; Andersen,~T.; Mcleod,~A.~S.; Fei,~Z.; Liu,~M.~K.; Wagner,~M.;
  Watanabe,~K.; Taniguchi,~T.; Thiemens,~M.; Keilmann,~F.; Jarillo-Herrero,~P.;
  Fogler,~M.~M.; Basov,~D.~N. Subdiffractional focusing and guiding of
  polaritonic rays in a natural hyperbolic material. \emph{Nature
  Communications} \textbf{2015}, \emph{6}, 6963\relax
\mciteBstWouldAddEndPuncttrue
\mciteSetBstMidEndSepPunct{\mcitedefaultmidpunct}
{\mcitedefaultendpunct}{\mcitedefaultseppunct}\relax
\EndOfBibitem
\bibitem[Autore \latin{et~al.}(2018)Autore, Li, Dolado, Alfaro-Mozaz, Esteban,
  Atxabal, Casanova, Hueso, Alonso-González, Aizpurua, Nikitin, Vélez, and
  Hillenbrand]{autore_boron_2018}
Autore,~M.; Li,~P.; Dolado,~I.; Alfaro-Mozaz,~F.~J.; Esteban,~R.; Atxabal,~A.;
  Casanova,~F.; Hueso,~L.~E.; Alonso-González,~P.; Aizpurua,~J.;
  Nikitin,~A.~Y.; Vélez,~S.; Hillenbrand,~R. Boron nitride nanoresonators for
  phonon-enhanced molecular vibrational spectroscopy at the strong coupling
  limit. \emph{Light: Science \& Applications} \textbf{2018}, \emph{7},
  17172--17172\relax
\mciteBstWouldAddEndPuncttrue
\mciteSetBstMidEndSepPunct{\mcitedefaultmidpunct}
{\mcitedefaultendpunct}{\mcitedefaultseppunct}\relax
\EndOfBibitem
\bibitem[Bareza \latin{et~al.}(2022)Bareza, Paulillo, Slipchenko, Autore,
  Dolado, Liu, Edgar, Vélez, Martín-Moreno, Hillenbrand, and
  Pruneri]{bareza_phonon-enhanced_2022}
Bareza,~N.~J.; Paulillo,~B.; Slipchenko,~T.~M.; Autore,~M.; Dolado,~I.;
  Liu,~S.; Edgar,~J.~H.; Vélez,~S.; Martín-Moreno,~L.; Hillenbrand,~R.;
  Pruneri,~V. Phonon-Enhanced Mid-Infrared {CO}2 Gas Sensing Using Boron
  Nitride Nanoresonators. \emph{{ACS} Photonics} \textbf{2022}, \emph{9},
  34--42\relax
\mciteBstWouldAddEndPuncttrue
\mciteSetBstMidEndSepPunct{\mcitedefaultmidpunct}
{\mcitedefaultendpunct}{\mcitedefaultseppunct}\relax
\EndOfBibitem
\bibitem[Zheng \latin{et~al.}(2022)Zheng, Sun, Xu, Huang, Chen, Ke, Zhan, Chen,
  and Deng]{zheng_tunable_2022}
Zheng,~Z.; Sun,~F.; Xu,~N.; Huang,~W.; Chen,~X.; Ke,~Y.; Zhan,~R.; Chen,~H.;
  Deng,~S. Tunable Hyperbolic Phonon Polaritons in a Suspended van der Waals
  $\alpha$-MoO$_3$ with Gradient Gaps. \emph{Advanced Optical Materials}
  \textbf{2022}, \emph{10}, 2102057\relax
\mciteBstWouldAddEndPuncttrue
\mciteSetBstMidEndSepPunct{\mcitedefaultmidpunct}
{\mcitedefaultendpunct}{\mcitedefaultseppunct}\relax
\EndOfBibitem
\bibitem[Álvarez Pérez \latin{et~al.}(2022)Álvarez Pérez, Duan,
  Taboada-Gutiérrez, Ou, Nikulina, Liu, Edgar, Bao, Giannini, Hillenbrand,
  Martín-Sánchez, Nikitin, and Alonso-González]{alvarez-perez_negative_2022}
Álvarez Pérez,~G.; Duan,~J.; Taboada-Gutiérrez,~J.; Ou,~Q.; Nikulina,~E.;
  Liu,~S.; Edgar,~J.~H.; Bao,~Q.; Giannini,~V.; Hillenbrand,~R.;
  Martín-Sánchez,~J.; Nikitin,~A.~Y.; Alonso-González,~P. Negative
  reflection of nanoscale-confined polaritons in a low-loss natural medium.
  \emph{Science Advances} \textbf{2022}, \emph{8}, eabp8486\relax
\mciteBstWouldAddEndPuncttrue
\mciteSetBstMidEndSepPunct{\mcitedefaultmidpunct}
{\mcitedefaultendpunct}{\mcitedefaultseppunct}\relax
\EndOfBibitem
\bibitem[Yu \latin{et~al.}(2023)Yu, Yao, Hu, Jiang, Zheng, Fan, Heinz, and
  Fan]{yu_hyperbolic_2023}
Yu,~S.-J.; Yao,~H.; Hu,~G.; Jiang,~Y.; Zheng,~X.; Fan,~S.; Heinz,~T.~F.;
  Fan,~J.~A. Hyperbolic Polaritonic Rulers Based on van der Waals
  $\alpha$-MoO$_3$ Waveguides and Resonators. \emph{{ACS} Nano} \textbf{2023},
  \emph{17}, 23057--23064\relax
\mciteBstWouldAddEndPuncttrue
\mciteSetBstMidEndSepPunct{\mcitedefaultmidpunct}
{\mcitedefaultendpunct}{\mcitedefaultseppunct}\relax
\EndOfBibitem
\bibitem[Hu \latin{et~al.}(2025)Hu, Ou, Si, Wu, Wu, Dai, Krasnok, Mazor, Zhang,
  Bao, Qiu, and Alù]{hu_topological_2020}
Hu,~G.; Ou,~Q.; Si,~G.; Wu,~Y.; Wu,~J.; Dai,~Z.; Krasnok,~A.; Mazor,~Y.;
  Zhang,~Q.; Bao,~Q.; Qiu,~C.-W.; Alù,~A. Topological polaritons and photonic
  magic angles in twisted $\alpha$-MoO$_3$ bilayers. \emph{Nature}
  \textbf{2025}, \emph{582}, 209--213\relax
\mciteBstWouldAddEndPuncttrue
\mciteSetBstMidEndSepPunct{\mcitedefaultmidpunct}
{\mcitedefaultendpunct}{\mcitedefaultseppunct}\relax
\EndOfBibitem
\bibitem[Dai \latin{et~al.}(2020)Dai, Hu, Si, Ou, Zhang, Balendhran, Rahman,
  Zhang, Ou, Li, Alù, Qiu, and Bao]{dai_edge-oriented_2020}
Dai,~Z.; Hu,~G.; Si,~G.; Ou,~Q.; Zhang,~Q.; Balendhran,~S.; Rahman,~F.;
  Zhang,~B.~Y.; Ou,~J.~Z.; Li,~G.; Alù,~A.; Qiu,~C.-W.; Bao,~Q. Edge-oriented
  and steerable hyperbolic polaritons in anisotropic van der Waals
  nanocavities. \emph{Nature Communications} \textbf{2020}, \emph{11},
  6086\relax
\mciteBstWouldAddEndPuncttrue
\mciteSetBstMidEndSepPunct{\mcitedefaultmidpunct}
{\mcitedefaultendpunct}{\mcitedefaultseppunct}\relax
\EndOfBibitem
\bibitem[Hu \latin{et~al.}(2023)Hu, Chen, Teng, Yu, Xue, Chen, Xiao, Qu, Hu,
  Chen, Sun, Li, de~Abajo, and Dai]{hu_gate-tunable_2023}
Hu,~H.; Chen,~N.; Teng,~H.; Yu,~R.; Xue,~M.; Chen,~K.; Xiao,~Y.; Qu,~Y.;
  Hu,~D.; Chen,~J.; Sun,~Z.; Li,~P.; de~Abajo,~F. J.~G.; Dai,~Q. Gate-tunable
  negative refraction of mid-infrared polaritons. \emph{Science} \textbf{2023},
  \emph{379}, 558--561\relax
\mciteBstWouldAddEndPuncttrue
\mciteSetBstMidEndSepPunct{\mcitedefaultmidpunct}
{\mcitedefaultendpunct}{\mcitedefaultseppunct}\relax
\EndOfBibitem
\bibitem[Teng \latin{et~al.}(2024)Teng, Chen, Hu, de~Abajo, and Dai]{Teng2024}
Teng,~H.; Chen,~N.; Hu,~H.; de~Abajo,~F. J.~G.; Dai,~Q. Steering and cloaking
  of hyperbolic polaritons at deep-subwavelength scales. \emph{Nature
  Communications} \textbf{2024}, \emph{15}, 1--6\relax
\mciteBstWouldAddEndPuncttrue
\mciteSetBstMidEndSepPunct{\mcitedefaultmidpunct}
{\mcitedefaultendpunct}{\mcitedefaultseppunct}\relax
\EndOfBibitem
\bibitem[Sakib \latin{et~al.}(2025)Sakib, Hussain, Stepanova, Harris,
  Bocanegra, Wu, Wickramasinghe, and
  Shcherbakov]{sakib2025vacancyengineeredphononpolaritonsvan}
Sakib,~M.~A.; Hussain,~N.; Stepanova,~M.; Harris,~W.; Bocanegra,~J.~J.; Wu,~R.;
  Wickramasinghe,~H.~K.; Shcherbakov,~M.~R. Vacancy-Engineered Phonon
  Polaritons in a van der Waals Crystal. 2025;
  \url{https://arxiv.org/abs/2309.05574}\relax
\mciteBstWouldAddEndPuncttrue
\mciteSetBstMidEndSepPunct{\mcitedefaultmidpunct}
{\mcitedefaultendpunct}{\mcitedefaultseppunct}\relax
\EndOfBibitem
\bibitem[Koski \latin{et~al.}(2012)Koski, Wessells, Reed, Cha, Kong, and
  Cui]{Koski2012}
Koski,~K.~J.; Wessells,~C.~D.; Reed,~B.~W.; Cha,~J.~J.; Kong,~D.; Cui,~Y.
  Chemical intercalation of zerovalent metals into 2D layered Bi 2Se 3
  nanoribbons. \emph{Journal of the American Chemical Society} \textbf{2012},
  \emph{134}, 13773--13779\relax
\mciteBstWouldAddEndPuncttrue
\mciteSetBstMidEndSepPunct{\mcitedefaultmidpunct}
{\mcitedefaultendpunct}{\mcitedefaultseppunct}\relax
\EndOfBibitem
\bibitem[Reed \latin{et~al.}(2019)Reed, Williams, Moser, and Koski]{Reed2019}
Reed,~B.~W.; Williams,~D.~R.; Moser,~B.~P.; Koski,~K.~J. Chemically Tuning
  Quantized Acoustic Phonons in 2D Layered MoO3 Nanoribbons. \emph{Nano
  Letters} \textbf{2019}, \emph{19}, 4406--4412\relax
\mciteBstWouldAddEndPuncttrue
\mciteSetBstMidEndSepPunct{\mcitedefaultmidpunct}
{\mcitedefaultendpunct}{\mcitedefaultseppunct}\relax
\EndOfBibitem
\bibitem[Reed \latin{et~al.}(2024)Reed, Chen, and Koski]{Reed2024}
Reed,~B.~W.; Chen,~E.; Koski,~K.~J. Chemochromism and Tunable Acoustic Phonons
  in Intercalated MoO3: Ag-, Bi-, In-, Mo-, Os-, Pd-, Pt-, Rh-, Ru-, Sb-, and
  W-MoO3. \emph{Nano Letters} \textbf{2024}, \emph{40}, 59\relax
\mciteBstWouldAddEndPuncttrue
\mciteSetBstMidEndSepPunct{\mcitedefaultmidpunct}
{\mcitedefaultendpunct}{\mcitedefaultseppunct}\relax
\EndOfBibitem
\bibitem[Reed \latin{et~al.}(2024)Reed, Chen, and Koski]{Reed2024+}
Reed,~B.~W.; Chen,~E.; Koski,~K.~J. Tunable Chemochromism and Elastic
  Properties in Intercalated MoO3: Au-, Cr-, Fe-, Ge-, Mn-, and Ni-MoO3.
  \emph{ACS Nano} \textbf{2024}, \emph{18}, 12845--12852\relax
\mciteBstWouldAddEndPuncttrue
\mciteSetBstMidEndSepPunct{\mcitedefaultmidpunct}
{\mcitedefaultendpunct}{\mcitedefaultseppunct}\relax
\EndOfBibitem
\bibitem[Wu \latin{et~al.}(2020)Wu, Ou, Yin, Li, Ma, Yu, Liu, Cui, Bao, Duan,
  Álvarez Pérez, Dai, Shabbir, Medhekar, Li, Li, Alonso-González, and
  Bao]{wu_chemical_2020}
Wu,~Y. \latin{et~al.}  Chemical switching of low-loss phonon polaritons in
  $\alpha$-MoO$_3$ by hydrogen intercalation. \emph{Nature Communications}
  \textbf{2020}, \emph{11}, 2646, Publisher: Nature Publishing Group\relax
\mciteBstWouldAddEndPuncttrue
\mciteSetBstMidEndSepPunct{\mcitedefaultmidpunct}
{\mcitedefaultendpunct}{\mcitedefaultseppunct}\relax
\EndOfBibitem
\bibitem[Zheng \latin{et~al.}(2018)Zheng, Chen, Wang, Wang, Chen, Liu, Xu, Xie,
  Chen, Deng, and Xu]{zheng_highly_2018}
Zheng,~Z.; Chen,~J.; Wang,~Y.; Wang,~X.; Chen,~X.; Liu,~P.; Xu,~J.; Xie,~W.;
  Chen,~H.; Deng,~S.; Xu,~N. Highly Confined and Tunable Hyperbolic Phonon
  Polaritons in Van Der Waals Semiconducting Transition Metal Oxides.
  \emph{Advanced Materials} \textbf{2018}, \emph{30}, 1705318\relax
\mciteBstWouldAddEndPuncttrue
\mciteSetBstMidEndSepPunct{\mcitedefaultmidpunct}
{\mcitedefaultendpunct}{\mcitedefaultseppunct}\relax
\EndOfBibitem
\bibitem[Wu \latin{et~al.}(2021)Wu, Ou, Dong, Hu, Si, Dai, Qiu, Fuhrer,
  Mokkapati, and Bao]{wu_efficient_2021}
Wu,~Y.; Ou,~Q.; Dong,~S.; Hu,~G.; Si,~G.; Dai,~Z.; Qiu,~C.-W.; Fuhrer,~M.~S.;
  Mokkapati,~S.; Bao,~Q. Efficient and Tunable Reflection of Phonon Polaritons
  at Built-In Intercalation Interfaces. \emph{Advanced Materials}
  \textbf{2021}, \emph{33}, 2008070\relax
\mciteBstWouldAddEndPuncttrue
\mciteSetBstMidEndSepPunct{\mcitedefaultmidpunct}
{\mcitedefaultendpunct}{\mcitedefaultseppunct}\relax
\EndOfBibitem
\bibitem[Zhao \latin{et~al.}(2022)Zhao, Chen, Xue, Chen, Jia, Chen, Bao, Gao,
  and Chen]{zhao_ultralow-loss_2022}
Zhao,~Y.; Chen,~J.; Xue,~M.; Chen,~R.; Jia,~S.; Chen,~J.; Bao,~L.; Gao,~H.-J.;
  Chen,~J. Ultralow-Loss Phonon Polaritons in the Isotope-Enriched
  $\alpha$-MoO$_3$. \emph{Nano Letters} \textbf{2022}, \emph{22},
  10208--10215\relax
\mciteBstWouldAddEndPuncttrue
\mciteSetBstMidEndSepPunct{\mcitedefaultmidpunct}
{\mcitedefaultendpunct}{\mcitedefaultseppunct}\relax
\EndOfBibitem
\bibitem[Taboada-Gutiérrez \latin{et~al.}(2020)Taboada-Gutiérrez, Álvarez
  Pérez, Duan, Ma, Crowley, Prieto, Bylinkin, Autore, Volkova, Kimura, Kimura,
  Berger, Li, Bao, Gao, Errea, Nikitin, Hillenbrand, Martín-Sánchez, and
  Alonso-González]{taboada-gutierrez_broad_2020}
Taboada-Gutiérrez,~J. \latin{et~al.}  Broad spectral tuning of ultra-low-loss
  polaritons in a van der Waals crystal by intercalation. \emph{Nature
  Materials} \textbf{2020}, \emph{19}, 964--968\relax
\mciteBstWouldAddEndPuncttrue
\mciteSetBstMidEndSepPunct{\mcitedefaultmidpunct}
{\mcitedefaultendpunct}{\mcitedefaultseppunct}\relax
\EndOfBibitem
\bibitem[Yang \latin{et~al.}(2019)Yang, Xiao, Ma, Cui, Zhang, Zhai, Meng, Wang,
  Wei, Du, Li, Sun, Yang, Zhang, and
  Gong]{https://doi.org/10.1002/aenm.201803137}
Yang,~W.; Xiao,~J.; Ma,~Y.; Cui,~S.; Zhang,~P.; Zhai,~P.; Meng,~L.; Wang,~X.;
  Wei,~Y.; Du,~Z.; Li,~B.; Sun,~Z.; Yang,~S.; Zhang,~Q.; Gong,~Y. Tin
  Intercalated Ultrathin MoO3 Nanoribbons for Advanced Lithium–Sulfur
  Batteries. \emph{Advanced Energy Materials} \textbf{2019}, \emph{9},
  1803137\relax
\mciteBstWouldAddEndPuncttrue
\mciteSetBstMidEndSepPunct{\mcitedefaultmidpunct}
{\mcitedefaultendpunct}{\mcitedefaultseppunct}\relax
\EndOfBibitem
\bibitem[Wu \latin{et~al.}(2018)Wu, Xie, Li, Liu, Ding, Tao, Chen, Liu, Chen,
  Chu, Zhang, and Song]{doi:10.1021/acs.jpclett.7b03374}
Wu,~C.; Xie,~H.; Li,~D.; Liu,~D.; Ding,~S.; Tao,~S.; Chen,~H.; Liu,~Q.;
  Chen,~S.; Chu,~W.; Zhang,~B.; Song,~L. Atomically Intercalating Tin Ions into
  the Interlayer of Molybdenum Oxide Nanobelt toward Long-Cycling Lithium
  Battery. \emph{The Journal of Physical Chemistry Letters} \textbf{2018},
  \emph{9}, 817--824\relax
\mciteBstWouldAddEndPuncttrue
\mciteSetBstMidEndSepPunct{\mcitedefaultmidpunct}
{\mcitedefaultendpunct}{\mcitedefaultseppunct}\relax
\EndOfBibitem
\bibitem[Xu \latin{et~al.}(2020)Xu, Lai, Xia, Luo, Chen, Wang, Chen, Wang, Shi,
  Xie, and Liu]{xu_electronic_2020}
Xu,~X.; Lai,~H.; Xia,~Y.; Luo,~T.; Chen,~Y.; Wang,~S.; Chen,~K.; Wang,~X.;
  Shi,~T.; Xie,~W.; Liu,~P. The electronic properties tuned by the synergy of
  polaron and d-orbital in a Co–Sn co-intercalated $\alpha$-MoO$_3$ system.
  \emph{Journal of Materials Chemistry C} \textbf{2020}, \emph{8},
  6536--6541\relax
\mciteBstWouldAddEndPuncttrue
\mciteSetBstMidEndSepPunct{\mcitedefaultmidpunct}
{\mcitedefaultendpunct}{\mcitedefaultseppunct}\relax
\EndOfBibitem
\bibitem[Shcherbakov \latin{et~al.}(2025)Shcherbakov, Potma, Sugawara, Nowak,
  Stepanova, Davies, Davies-Jones, and
  Wickramasinghe]{shcherbakov_photo-induced_2025}
Shcherbakov,~M.~R.; Potma,~E.~O.; Sugawara,~Y.; Nowak,~D.; Stepanova,~M.;
  Davies,~P.~R.; Davies-Jones,~J.; Wickramasinghe,~H.~K. Photo-induced force
  microscopy. \emph{Nature Reviews Methods Primers} \textbf{2025}, \emph{5},
  1--16\relax
\mciteBstWouldAddEndPuncttrue
\mciteSetBstMidEndSepPunct{\mcitedefaultmidpunct}
{\mcitedefaultendpunct}{\mcitedefaultseppunct}\relax
\EndOfBibitem
\bibitem[Álvarez Pérez \latin{et~al.}(2020)Álvarez Pérez, Folland, Errea,
  Taboada-Gutiérrez, Duan, Martín-Sánchez, Tresguerres-Mata, Matson,
  Bylinkin, He, Ma, Bao, Martín, Caldwell, Nikitin, and
  Alonso-González]{alvarez-perez_infrared_2020}
Álvarez Pérez,~G. \latin{et~al.}  Infrared Permittivity of the Biaxial van
  der Waals Semiconductor $\alpha$-MoO$_3$ from Near- and Far-Field Correlative
  Studies. \emph{Advanced Materials} \textbf{2020}, \emph{32}, 1908176\relax
\mciteBstWouldAddEndPuncttrue
\mciteSetBstMidEndSepPunct{\mcitedefaultmidpunct}
{\mcitedefaultendpunct}{\mcitedefaultseppunct}\relax
\EndOfBibitem
\bibitem[Ma \latin{et~al.}(2018)Ma, Alonso-González, Li, Nikitin, Yuan,
  Martín-Sánchez, Taboada-Gutiérrez, Amenabar, Li, Vélez, Tollan, Dai,
  Zhang, Sriram, Kalantar-Zadeh, Lee, Hillenbrand, and Bao]{ma_-plane_2018}
Ma,~W. \latin{et~al.}  In-plane anisotropic and ultra-low-loss polaritons in a
  natural van der Waals crystal. \emph{Nature} \textbf{2018}, \emph{562},
  557--562\relax
\mciteBstWouldAddEndPuncttrue
\mciteSetBstMidEndSepPunct{\mcitedefaultmidpunct}
{\mcitedefaultendpunct}{\mcitedefaultseppunct}\relax
\EndOfBibitem
\bibitem[Huang \latin{et~al.}(2014)Huang, He, Cao, and Lu]{huang_impact_2014}
Huang,~P.-R.; He,~Y.; Cao,~C.; Lu,~Z.-H. Impact of lattice distortion and
  electron doping on $\alpha$-MoO$_3$ electronic structure. \emph{Scientific
  Reports} \textbf{2014}, \emph{4}, 7131\relax
\mciteBstWouldAddEndPuncttrue
\mciteSetBstMidEndSepPunct{\mcitedefaultmidpunct}
{\mcitedefaultendpunct}{\mcitedefaultseppunct}\relax
\EndOfBibitem
\bibitem[Rajapakse \latin{et~al.}(2021)Rajapakse, Karki, Abu, Pishgar, Musa,
  Riyadh, Yu, Sumanasekera, and Jasinski]{rajapakse_intercalation_2021}
Rajapakse,~M.; Karki,~B.; Abu,~U.~O.; Pishgar,~S.; Musa,~M. R.~K.; Riyadh,~S.
  M.~S.; Yu,~M.; Sumanasekera,~G.; Jasinski,~J.~B. Intercalation as a versatile
  tool for fabrication, property tuning, and phase transitions in 2D materials.
  \emph{npj 2D Materials and Applications} \textbf{2021}, \emph{5}, 30\relax
\mciteBstWouldAddEndPuncttrue
\mciteSetBstMidEndSepPunct{\mcitedefaultmidpunct}
{\mcitedefaultendpunct}{\mcitedefaultseppunct}\relax
\EndOfBibitem
\bibitem[Quan \latin{et~al.}(2020)Quan, Zhang, Wei, Li, Park, Hwang, Tian,
  Huang, Wang, Wang, Kwak, Qin, Peng, and Ruoff]{doi:10.1021/acsami.0c02301}
Quan,~L.; Zhang,~H.; Wei,~H.; Li,~Y.; Park,~S.~O.; Hwang,~D.~Y.; Tian,~Y.;
  Huang,~M.; Wang,~C.; Wang,~M.; Kwak,~S.~K.; Qin,~F.; Peng,~H.-X.;
  Ruoff,~R.~S. The Electromagnetic Absorption of a Na-Ethylenediamine Graphite
  Intercalation Compound. \emph{ACS Applied Materials \& Interfaces}
  \textbf{2020}, \emph{12}, 16841--16848\relax
\mciteBstWouldAddEndPuncttrue
\mciteSetBstMidEndSepPunct{\mcitedefaultmidpunct}
{\mcitedefaultendpunct}{\mcitedefaultseppunct}\relax
\EndOfBibitem
\bibitem[Peelaers \latin{et~al.}(2017)Peelaers, Chabinyc, and Van~de
  Walle]{peelaers_controlling_2017}
Peelaers,~H.; Chabinyc,~M.~L.; Van~de Walle,~C.~G. Controlling n-Type Doping in
  {MoO}3. \emph{Chemistry of Materials} \textbf{2017}, \emph{29},
  2563--2567\relax
\mciteBstWouldAddEndPuncttrue
\mciteSetBstMidEndSepPunct{\mcitedefaultmidpunct}
{\mcitedefaultendpunct}{\mcitedefaultseppunct}\relax
\EndOfBibitem
\bibitem[de~Oliveira~Campos \latin{et~al.}(2025)de~Oliveira~Campos, Barbosa,
  de~Morais, Melo, de~Jesus~Santana, and
  Patience]{https://doi.org/10.1002/cjce.25712}
de~Oliveira~Campos,~V.; Barbosa,~F.~F.; de~Morais,~E. K.~L.; Melo,~D. M.~A.;
  de~Jesus~Santana,~J.; Patience,~G.~S. Experimental methods in chemical
  engineering: Electron probe micro-analysis—EPMA. \emph{The Canadian Journal
  of Chemical Engineering} \textbf{2025}, \emph{103}, 3000--3011\relax
\mciteBstWouldAddEndPuncttrue
\mciteSetBstMidEndSepPunct{\mcitedefaultmidpunct}
{\mcitedefaultendpunct}{\mcitedefaultseppunct}\relax
\EndOfBibitem
\bibitem[Matthews \latin{et~al.}(2021)Matthews, Kearns, and
  Buse]{10.1017/S1431927621000192}
Matthews,~M.~B.; Kearns,~S.~L.; Buse,~B. Low-Voltage Electron-Probe
  Microanalysis of Uranium. \emph{Microscopy and Microanalysis} \textbf{2021},
  \emph{27}, 466--483\relax
\mciteBstWouldAddEndPuncttrue
\mciteSetBstMidEndSepPunct{\mcitedefaultmidpunct}
{\mcitedefaultendpunct}{\mcitedefaultseppunct}\relax
\EndOfBibitem
\bibitem[Magnussen \latin{et~al.}(2024)Magnussen, Drnec, Qiu, Martens, Huang,
  Chattot, and Singer]{doi:10.1021/acs.chemrev.3c00331}
Magnussen,~O.~M.; Drnec,~J.; Qiu,~C.; Martens,~I.; Huang,~J.~J.; Chattot,~R.;
  Singer,~A. In Situ and Operando X-ray Scattering Methods in Electrochemistry
  and Electrocatalysis. \emph{Chemical Reviews} \textbf{2024}, \emph{124},
  629--721, PMID: 38253355\relax
\mciteBstWouldAddEndPuncttrue
\mciteSetBstMidEndSepPunct{\mcitedefaultmidpunct}
{\mcitedefaultendpunct}{\mcitedefaultseppunct}\relax
\EndOfBibitem
\bibitem[Liu \latin{et~al.}(2025)Liu, Xiong, Wang, Bai, Ma, Wang, Li, Li, Wang,
  Garcia-Vidal, Dai, and Hu]{liu_long-range_2025}
Liu,~L.; Xiong,~L.; Wang,~C.; Bai,~Y.; Ma,~W.; Wang,~Y.; Li,~P.; Li,~G.;
  Wang,~Q.~J.; Garcia-Vidal,~F.~J.; Dai,~Z.; Hu,~G. Long-range hyperbolic
  polaritons on a non-hyperbolic crystal surface. \emph{Nature} \textbf{2025},
  1--7\relax
\mciteBstWouldAddEndPuncttrue
\mciteSetBstMidEndSepPunct{\mcitedefaultmidpunct}
{\mcitedefaultendpunct}{\mcitedefaultseppunct}\relax
\EndOfBibitem
\bibitem[Hu \latin{et~al.}(2023)Hu, Ma, Hu, Wu, Zheng, Liu, Zhang, Ni, Chen,
  Zhang, Dai, Caldwell, Paarmann, Alù, Li, and Qiu]{hu_real-space_2023}
Hu,~G. \latin{et~al.}  Real-space nanoimaging of hyperbolic shear polaritons in
  a monoclinic crystal. \emph{Nature Nanotechnology} \textbf{2023}, \emph{18},
  64--70, Publisher: Nature Publishing Group\relax
\mciteBstWouldAddEndPuncttrue
\mciteSetBstMidEndSepPunct{\mcitedefaultmidpunct}
{\mcitedefaultendpunct}{\mcitedefaultseppunct}\relax
\EndOfBibitem
\bibitem[Passler \latin{et~al.}(2022)Passler, Ni, Hu, Matson, Carini, Wolf,
  Schubert, Alù, Caldwell, Folland, and Paarmann]{passler_hyperbolic_2022}
Passler,~N.~C.; Ni,~X.; Hu,~G.; Matson,~J.~R.; Carini,~G.; Wolf,~M.;
  Schubert,~M.; Alù,~A.; Caldwell,~J.~D.; Folland,~T.~G.; Paarmann,~A.
  Hyperbolic shear polaritons in low-symmetry crystals. \emph{Nature}
  \textbf{2022}, \emph{602}, 595--600, Publisher: Nature Publishing Group\relax
\mciteBstWouldAddEndPuncttrue
\mciteSetBstMidEndSepPunct{\mcitedefaultmidpunct}
{\mcitedefaultendpunct}{\mcitedefaultseppunct}\relax
\EndOfBibitem
\bibitem[Li \latin{et~al.}(2006)Li, Jiang, Pang, Peng, and Zhang]{Li2006}
Li,~G.; Jiang,~L.; Pang,~S.; Peng,~H.; Zhang,~Z. Molybdenum trioxide
  nanostructures: The evolution from helical nanosheets to crosslike
  nanoflowers to nanobelts. \emph{Journal of Physical Chemistry B}
  \textbf{2006}, \emph{110}, 24472--24475\relax
\mciteBstWouldAddEndPuncttrue
\mciteSetBstMidEndSepPunct{\mcitedefaultmidpunct}
{\mcitedefaultendpunct}{\mcitedefaultseppunct}\relax
\EndOfBibitem
\bibitem[Wang and Koski(2015)Wang, and Koski]{Wang2015}
Wang,~M.; Koski,~K.~J. Reversible chemochromic MoO3 nanoribbons through
  zerovalent metal intercalation. \emph{ACS Nano} \textbf{2015}, \emph{9},
  3226--3233\relax
\mciteBstWouldAddEndPuncttrue
\mciteSetBstMidEndSepPunct{\mcitedefaultmidpunct}
{\mcitedefaultendpunct}{\mcitedefaultseppunct}\relax
\EndOfBibitem
\bibitem[Kresse and Furthm{\"u}ller(1996)Kresse, and
  Furthm{\"u}ller]{kresse1996efficient}
Kresse,~G.; Furthm{\"u}ller,~J. Efficient iterative schemes for \textit{ab
  initio} total-energy calculations using a plane-wave basis set.
  \emph{Physical Review B} \textbf{1996}, \emph{54}, 11169--11186\relax
\mciteBstWouldAddEndPuncttrue
\mciteSetBstMidEndSepPunct{\mcitedefaultmidpunct}
{\mcitedefaultendpunct}{\mcitedefaultseppunct}\relax
\EndOfBibitem
\bibitem[Kresse and Hafner(1993)Kresse, and Hafner]{kresse1993ab}
Kresse,~G.; Hafner,~J. \textit{ab initio} molecular dynamics for liquid metals.
  \emph{Physical Review B} \textbf{1993}, \emph{47}, 558--561\relax
\mciteBstWouldAddEndPuncttrue
\mciteSetBstMidEndSepPunct{\mcitedefaultmidpunct}
{\mcitedefaultendpunct}{\mcitedefaultseppunct}\relax
\EndOfBibitem
\bibitem[Grimme \latin{et~al.}(2010)Grimme, Antony, Ehrlich, and
  Krieg]{grimme2011effect}
Grimme,~S.; Antony,~J.; Ehrlich,~S.; Krieg,~H. A consistent and accurate ab
  initio parametrization of density functional dispersion correction (DFT-D)
  for the 94 elements H-Pu. \emph{The Journal of Chemical Physics}
  \textbf{2010}, \emph{132}, 154104\relax
\mciteBstWouldAddEndPuncttrue
\mciteSetBstMidEndSepPunct{\mcitedefaultmidpunct}
{\mcitedefaultendpunct}{\mcitedefaultseppunct}\relax
\EndOfBibitem
\bibitem[Monkhorst and Pack(1976)Monkhorst, and Pack]{monkhorst1976special}
Monkhorst,~H.~J.; Pack,~J.~D. Special points for Brillouin-zone integrations.
  \emph{Physical Review B} \textbf{1976}, \emph{13}, 5188--5192\relax
\mciteBstWouldAddEndPuncttrue
\mciteSetBstMidEndSepPunct{\mcitedefaultmidpunct}
{\mcitedefaultendpunct}{\mcitedefaultseppunct}\relax
\EndOfBibitem
\bibitem[Langreth \latin{et~al.}(2005)Langreth, Dion, Rydberg, Schr{\"o}der,
  Hyldgaard, and Lundqvist]{langreth2005van}
Langreth,~D.~C.; Dion,~M.; Rydberg,~H.; Schr{\"o}der,~E.; Hyldgaard,~P.;
  Lundqvist,~B.~I. Van der Waals density functional theory with applications.
  \emph{International Journal of Quantum Chemistry} \textbf{2005}, \emph{101},
  599--610\relax
\mciteBstWouldAddEndPuncttrue
\mciteSetBstMidEndSepPunct{\mcitedefaultmidpunct}
{\mcitedefaultendpunct}{\mcitedefaultseppunct}\relax
\EndOfBibitem
\bibitem[Tong \latin{et~al.}(2021)Tong, Dumitric{\u{a}}, and
  Frauenheim]{tong2021first}
Tong,~Z.; Dumitric{\u{a}},~T.; Frauenheim,~T. First-principles prediction of
  infrared phonon and dielectric function in biaxial hyperbolic van der Waals
  crystal $\alpha$-MoO$_3$. \emph{Physical Chemistry Chemical Physics}
  \textbf{2021}, \emph{23}, 19627--19635\relax
\mciteBstWouldAddEndPuncttrue
\mciteSetBstMidEndSepPunct{\mcitedefaultmidpunct}
{\mcitedefaultendpunct}{\mcitedefaultseppunct}\relax
\EndOfBibitem
\bibitem[Zhao \latin{et~al.}(2022)Zhao, Chen, Xue, Chen, Jia, Chen, Bao, Gao,
  and Chen]{zhao2022ultralow}
Zhao,~Y.; Chen,~J.; Xue,~M.; Chen,~R.; Jia,~S.; Chen,~J.; Bao,~L.; Gao,~H.-J.;
  Chen,~J. Ultralow-Loss Phonon Polaritons in the Isotope-Enriched
  $\alpha$-MoO3. \emph{Nano Letters} \textbf{2022}, \emph{22},
  10208--10215\relax
\mciteBstWouldAddEndPuncttrue
\mciteSetBstMidEndSepPunct{\mcitedefaultmidpunct}
{\mcitedefaultendpunct}{\mcitedefaultseppunct}\relax
\EndOfBibitem
\bibitem[Taboada-Guti{\'e}rrez \latin{et~al.}(2024)Taboada-Guti{\'e}rrez, Zhou,
  Tresguerres-Mata, Lanza, Martínez-Suárez, Álvarez P{\'e}rez, Duan,
  Martín, V{\'e}lez, Prieto, Bercher, Teyssier, Errea, Nikitin,
  Martín-Sánchez, Kuzmenko, and
  Alonso-González]{taboada2024unveiling}
Taboada-Guti{\'e}rrez,~J. \latin{et~al.}  Unveiling the Mechanism of
  Phonon-Polariton Damping in $\alpha$-MoO3. \emph{ACS Photonics}
  \textbf{2024}, \emph{11}, 3570--3577\relax
\mciteBstWouldAddEndPuncttrue
\mciteSetBstMidEndSepPunct{\mcitedefaultmidpunct}
{\mcitedefaultendpunct}{\mcitedefaultseppunct}\relax
\EndOfBibitem
\bibitem[Zhou \latin{et~al.}(2023)Zhou, Song, Xu, Ni, Dang, Zhao, Quan, Dong,
  Hu, Huang, Chen, Wang, Cheng, Raschke, Alù, and
  Jiang]{zhou2023gate}
Zhou,~Z. \latin{et~al.}  Gate-Tuning Hybrid Polaritons in Twisted
  $\alpha$-MoO$_3$/Graphene Heterostructures. \emph{Nano Letters}
  \textbf{2023}, \emph{23}, 11252--11259\relax
\mciteBstWouldAddEndPuncttrue
\mciteSetBstMidEndSepPunct{\mcitedefaultmidpunct}
{\mcitedefaultendpunct}{\mcitedefaultseppunct}\relax
\EndOfBibitem
\end{mcitethebibliography}

\providecommand{\latin}[1]{#1}
\makeatletter
\providecommand{\doi}
  {\begingroup\let\do\@makeother\dospecials
  \catcode`\{=1 \catcode`\}=2 \doi@aux}
\providecommand{\doi@aux}[1]{\endgroup\texttt{#1}}
\makeatother
\providecommand*\mcitethebibliography{\thebibliography}
\csname @ifundefined\endcsname{endmcitethebibliography}
  {\let\endmcitethebibliography\endthebibliography}{}

\clearpage
\section{SUPPLEMENTARY INFORMATION}

\setcounter{figure}{0}

\renewcommand{\figurename}{Figure S}
\renewcommand{\tablename}{Table S}

\begin{figure}[h!]
  \includegraphics[height=6cm]{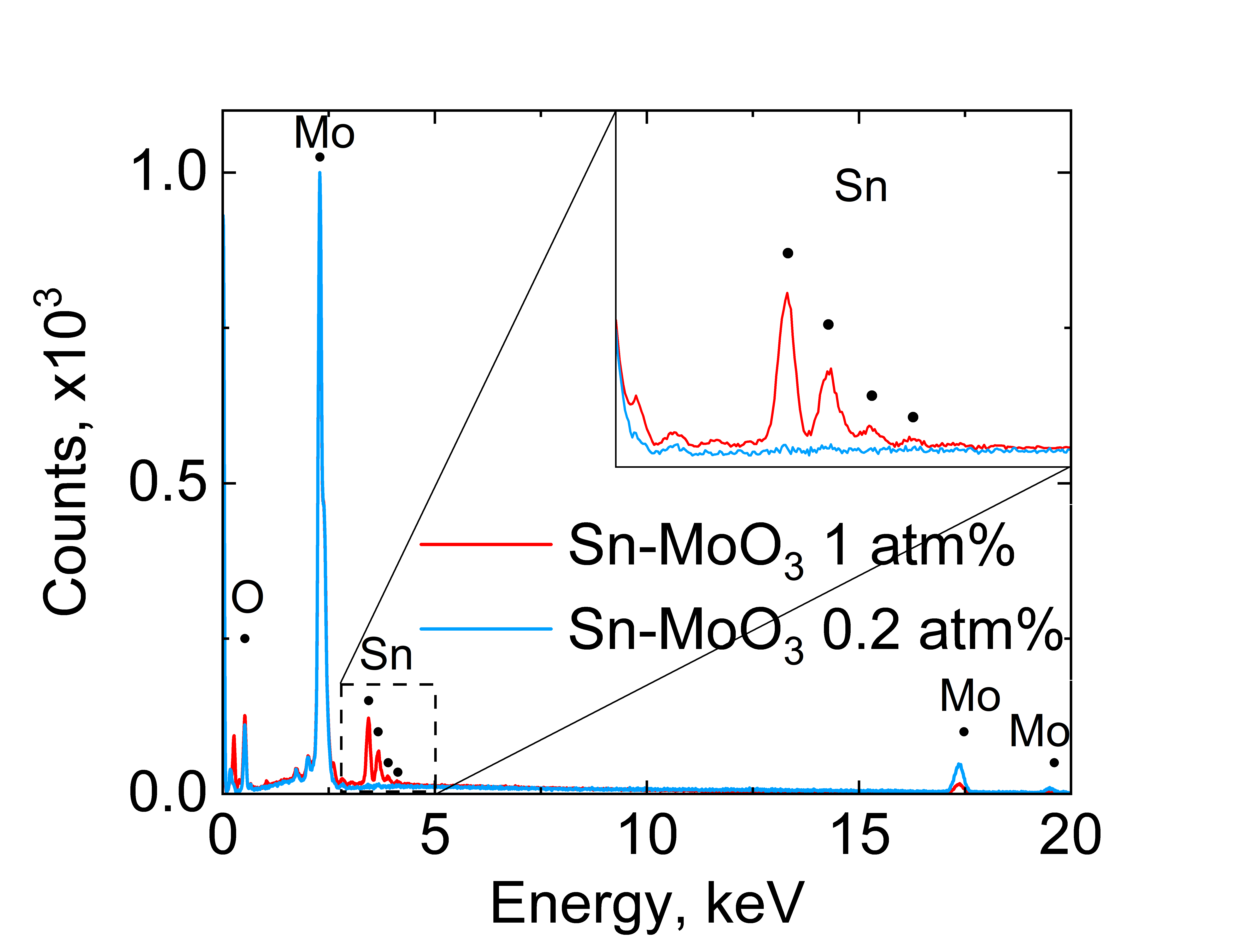}
  \caption{EDX spectra of 1 atm\% Sn-MoO$_3$ (red) and 115-nm-thick 0.2 atm\% Sn-MoO$_3$ (blue).} 
  \label{fig1}
\end{figure}

\begin{figure}[b!] 
  \includegraphics[height=10cm]{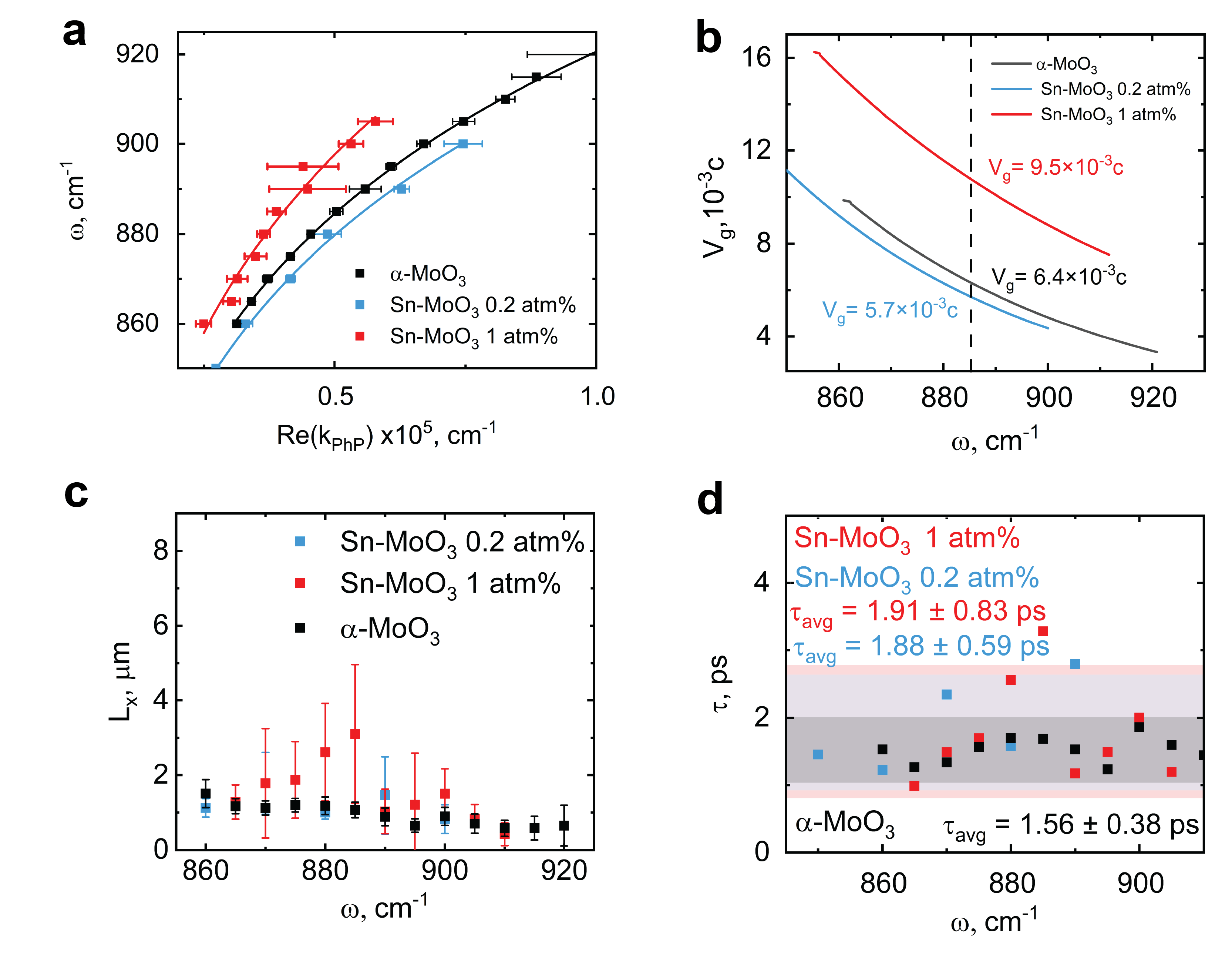}
  \caption{\textbf{a}, PhP dispersion, \textbf{b}, group velocity, \textbf{c}, propagation length, \textbf{d}, lifetime of 120-nm-thick $\alpha$-MoO$_3$ (black), 1 atm\% Sn-MoO$_3$ (red) and 115-nm-thick 0.2 atm\% Sn-MoO$_3$ (blue).} 
  \label{fig1}
\end{figure}

\begin{figure}
  \includegraphics[height=10cm]{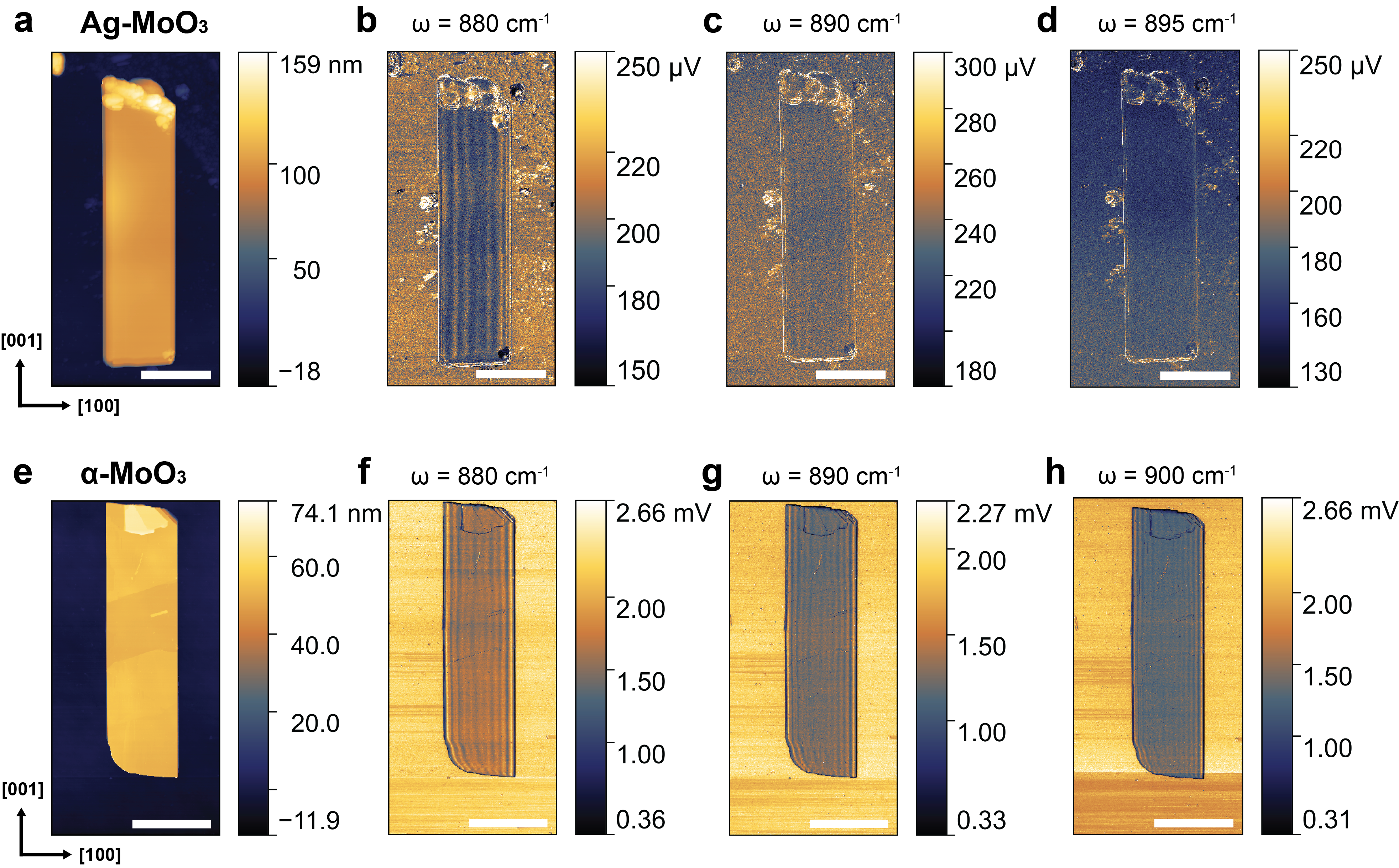}
  \caption{\textbf{a-d}, Topography and PiFM images of 97-nm-thick Ag-MoO$_3$ at 880-895 cm$^{-1}$ excitation. \textbf{e-h} Topography and PiFM images of 55 nm-thick $\alpha$-MoO$_3$ at 880-900 cm$^{-1}$ excitation. Scale bars are 3 $\mu$m.} 
  \label{fig1}
\end{figure}


\begin{figure}
  \includegraphics[height=4.5cm]{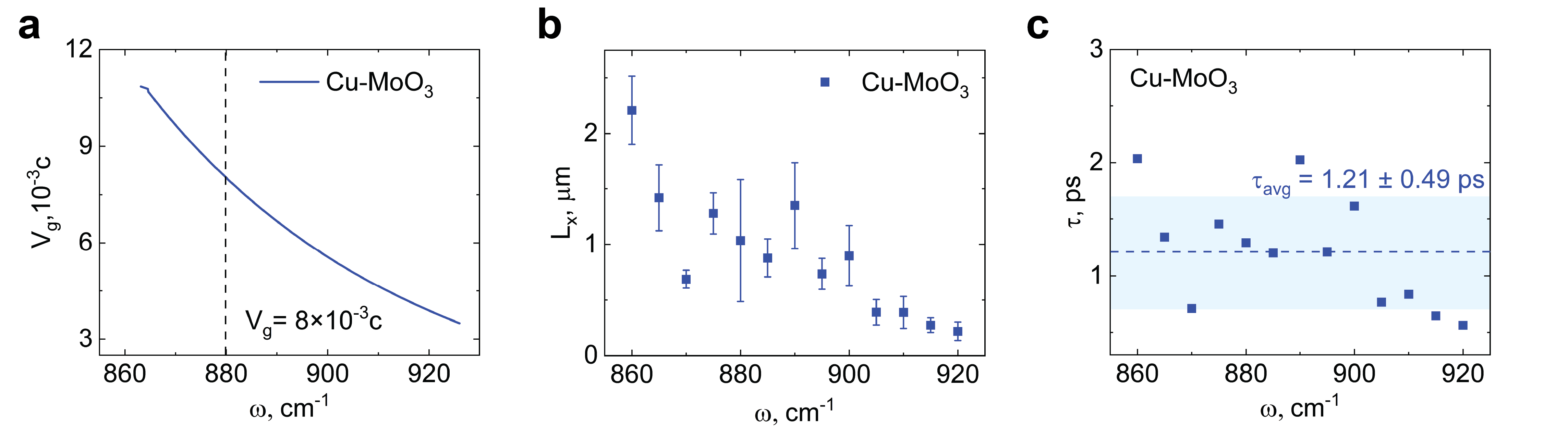}
  \caption{\textbf{a}, PhP group velocity, \textbf{b}, propagation length, \textbf{c}, lifetime of 140-nm-thick Cu-MoO$_3$.} 
  \label{fig1}
\end{figure}

\clearpage
\section{Supplementary Note S1: Analytical Model of PhP Dispersion}

To quantitatively estimate the shift of PhP dispersion induced by the modulation of static dielectric permittivity, an analytical model of PhP dispersion\cite{alvarez-perez_infrared_2020} is considered:

\begin{equation}
q=\frac{\rho}{k_0d}\left[\text{arctan}\left(\frac{\varepsilon_1\rho}{\varepsilon_z}\right)+\text{arctan}\left(\frac{\varepsilon_1\rho}{\varepsilon_z}\right)+\pi l \right], l=0,1,2...
\label{eq2}
\tag{S1}
\end{equation}

where \(q=k_t/k_0\) is the normalized in-plane momentum \((k_t^2=k_x^2+k_y^2)\), $\varepsilon_1$ and $\varepsilon_3$  are the permittivities of the superstrate (air) and substrate (silicon), respectively; $d$ is the thickness of the $\alpha$-MoO$_3$ flake, $k_0 = \omega/c$ is the wavevector in free space and \(\rho=i\sqrt{\varepsilon_z/(\varepsilon_x \text{cos}^2\phi+\varepsilon_y\text{sin}^2\phi)}\), with $\phi$ being the angle between the $x$ axis and the in-plane component vector. For 
$\omega= 820-970$~cm$^{-1}$
, where PhPs propagate 
along [001], 
$\phi=0^\circ$ and 
\(\rho=i\sqrt{\varepsilon_z/\varepsilon_x}\), making 
the PhP dispersion depend on dielectric permittivities $\varepsilon_x$ and $\varepsilon_z$. $\alpha$-MoO$_3$ is a biaxial crystal with the dielectric permittivity tensor $\hat{\varepsilon}(\omega)$:

\begin{equation}
\hat{\varepsilon}(\omega)= \begin{pmatrix}
\varepsilon_x(\omega) & 0 & 0\\
0 & \varepsilon_y(\omega) & 0\\
0 & 0 & \varepsilon_z(\omega)
\end{pmatrix}
\label{eq3}
\tag{S2}
\end{equation}

where all three principal values $\varepsilon_x(\omega)$, $\varepsilon_y(\omega)$, and $\varepsilon_z(\omega)$ are different and can be approximated by the Lorentz model\cite{alvarez-perez_infrared_2020} for coupled oscillators:

\begin{equation}
\varepsilon_j(\omega)=\varepsilon_\infty \left(1+\sum_j\frac{\omega_{LO}^2-\omega_{TO}^2}{\omega_{TO}^2-\omega^2-i\gamma\omega}\right), j=x,y,z
\label{eq4}
\tag{S3}
\end{equation}

where $x$, $y$ and $z$ correspond to the [100], [001], and [010] crystal directions, respectively. 

\begin{figure}
  \includegraphics[height=7.25cm]{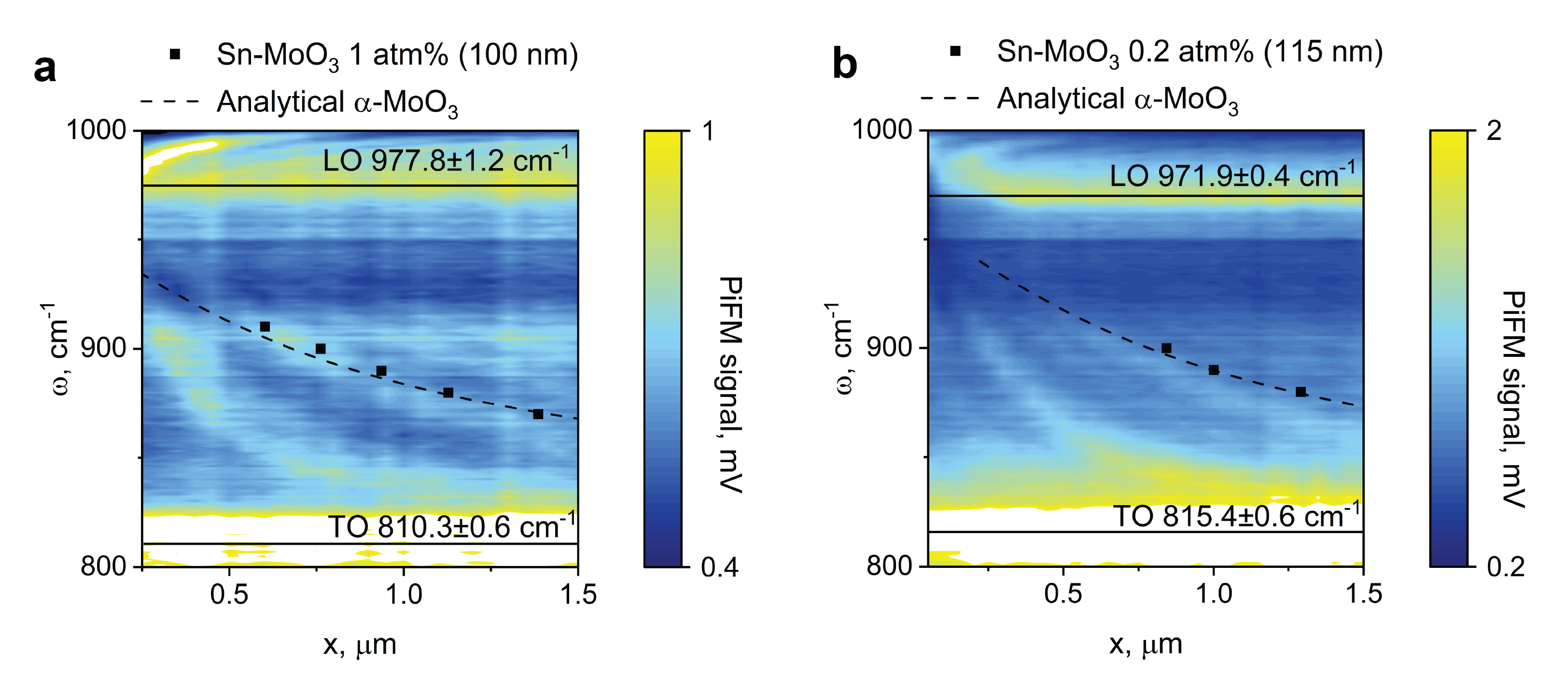}
  \caption{\textbf{a}, Spectral map along the $x$ [100] crystallographic direction of 100-nm-thick 1 atm\% Sn-MoO$_3$. \textbf{b}, Spectral map along the $x$ [100] crystallographic direction of 115-nm-thick 0.2 atm\% Sn-MoO$_3$.} 
  \label{fig1}
\end{figure}


\begin{table}
    \centering
    \begin{tabular}{ | m{10em} | m{4.5cm}| m{4cm} | m{3cm} | } 
        \hline
        Tuning type & Maximum $\Delta k/k $ 
        & Average $\Delta k/k $
        & Reference\\
        \hline
        Isotope & 11.8\% @ 875 cm$^{-1}$ & $7.07\pm0.03\%$ & \citeauthor{zhao2022ultralow}\cite{zhao2022ultralow}\\
        Thermal & 20.7\% @ 860 cm$^{-1}$ & $12.54\pm0.04\%$ & \citeauthor{taboada2024unveiling} \cite{taboada2024unveiling}\\
        Oxygen-vacancy & 17.0\% @ 870 cm$^{-1}$ & $15.4\pm1.5\%$ & \citeauthor{sakib2025vacancyengineeredphononpolaritonsvan}\cite{sakib2025vacancyengineeredphononpolaritonsvan}\\
         Gating (70 V) 
         & 28.6\% @ 931 cm$^{-1}$ & $22.4\pm6.1\%$ & \citeauthor{zhou2023gate}\cite{zhou2023gate}\\
        Gating (150 V) 
        & 30.5\% @ 893cm$^{-1}$ & $19.48\pm0.11\%$ & \citeauthor{hu_gate-tunable_2023}\cite{hu_gate-tunable_2023}\\
         Tin intercalation & $38.5\pm0.5\%$ @ 910 cm$^{-1}$ & $23.5\pm7.2\%$ & This work\\
        \hline
    \end{tabular}
    \caption{Comparison of non-mechanical PhP dispersion modulation approaches in $\alpha$-MoO$_3$}
    \label{tab:placeholder}
\end{table}

\end{document}